\documentclass[11pt,english]{extarticle}
\usepackage[T1]{fontenc}
\usepackage[latin9]{inputenc}
\usepackage{geometry}
\usepackage{color}
\usepackage{array}
\usepackage{enumitem}
\usepackage{multirow}
\usepackage{amsmath}
\usepackage{amsthm}
\usepackage{amssymb}
\usepackage{graphicx}
\usepackage{setspace}
\usepackage[numbers]{natbib}
\makeatletter

\providecommand{\tabularnewline}{\\}

\numberwithin{equation}{section}
\numberwithin{figure}{section}
\newcommand{\lyxaddress}[1]{
\par {\raggedright #1
\vspace{1.4em}
\noindent\par}
}
\theoremstyle{plain}
\newtheorem{thm}{\protect\theoremname}
\theoremstyle{plain}
\newtheorem{lem}[thm]{\protect\lemmaname}
\theoremstyle{remark}
\newtheorem{rem}[thm]{\protect\remarkname}
\theoremstyle{definition}
\newtheorem{defn}[thm]{\protect\definitionname}
\ifx\proof\undefined
\newenvironment{proof}[1][\protect\proofname]{\par
\normalfont\topsep6\p@\@plus6\p@\relax
\trivlist
\itemindent\parindent
\item[\hskip\labelsep\scshape #1]\ignorespaces
}{%
\endtrivlist\@endpefalse
}
\providecommand{\proofname}{Proof}
\fi
 \newlist{casenv}{enumerate}{4}
 \setlist[casenv]{leftmargin=*,align=left,widest={iiii}}
 \setlist[casenv,1]{label={{\itshape\ \casename} \arabic*.},ref=\arabic*}
 \setlist[casenv,2]{label={{\itshape\ \casename} \roman*.},ref=\roman*}
 \setlist[casenv,3]{label={{\itshape\ \casename\ \alph*.}},ref=\alph*}
 \setlist[casenv,4]{label={{\itshape\ \casename} \arabic*.},ref=\arabic*}

\@ifundefined{showcaptionsetup}{}{%
 \PassOptionsToPackage{caption=false}{subfig}}
\usepackage{subfig}
\makeatother

\usepackage{babel}
 \providecommand{\casename}{Case}
\providecommand{\definitionname}{Definition}
\providecommand{\lemmaname}{Lemma}
\providecommand{\remarkname}{Remark}
\providecommand{\theoremname}{Theorem}

\begin{document}

\title{Communication Melting in Graphs and Complex Networks}

\author{Najlaa Alalwan$^{1}$, Alex Arenas$^{2}$ and Ernesto Estrada$^{1}$}
\maketitle

\lyxaddress{$^{1}$Department of Mathematics and Statistics, University of Strathclyde,
26 Richmond Street, Glasgow G11HX, UK. $^{2}$Departament d'Enginyeria
Informàtica i Matemàtiques, Universitat Rovira i Virgili, 43007 Tarragona,
Spain. }
\begin{abstract}
Complex networks are the representative graphs of interactions in
many complex systems. Usually, these interactions are abstractions
of the communication/diffusion channels between the units of the system.
Real complex networks, e.g. traffic networks, reveal different operation
phases governed by the dynamical stress of the system. In the case
of traffic networks the archetypical transition is from free flow
to congestion. A revolutionary approach to ascertain how these transitions
emerge is that of using physical models that could account for diffusion
process under stress. Here we show how, communicability, a topological
descriptor that reveals the efficiency of the network functionality
in terms of these diffusive paths, could be used to reveal the transitions
mentioned. By considering a vibrational model of nodes and edges in
a graph/network at a given temperature (stress), we show that the
communicability function plays the role of the thermal Green's function
of a network of harmonic oscillators. After, we prove analytically
the existence of a universal phase transition in the communicability
structure of every simple graph. This transition resembles the melting
process occurring in solids. For instance, regular-like graphs resembling
crystals, melts at lower temperatures and display a sharper transition
between connected to disconnected structures than the random spatial
graphs, which resemble amorphous solids. Finally, we study computationally
this graph melting process in some real-world networks and observe
that the rate of melting of graphs changes either as an exponential
or as a power-law with the inverse temperature. At the local level
we discover that the main driver for node melting is the eigenvector
centrality of the corresponding node, particularly when the critical
value of the inverse temperature approaches zero. These universal
results sheds light on many dynamical diffusive-like processes on
networks that present transitions as traffic jams, communication lost
or failure cascades.
\end{abstract}

\section{Introduction}

The use of graphs and networks to represent many physical, biological,
social and engineering systems has triggered their relevance as an
object of study in applied mathematics \citep{estrada2012structure,estrada2013graph,newman2003structure}.
One of the main goals of these studies is to understand the robustness
of these networks to the external stresses to which they are constantly
submitted to. In this sense, the study of melting processes of graphs
and networks can bring some new lights on this important area of applied
research. A network can be considered as a general system of balls
and springs submerged into a thermal bath at a given inverse temperature
$\beta=\left(k_{B}T\right)^{-1}$ where $k_{B}$ is a constant \citep{estrada2012physics}.
Here the thermal bath represents the external stress to which the
system is submitted to and $\beta$ represents a weight applied to
every edge of the graph (see Preliminaries for details). When the
external stress is too strong, $\beta\rightarrow0$, the graph is
fully disconnected indicating that no transfer of ``information''
is possible between any pair of nodes in the network. The capacity
of a node to transmit a perturbation at a given $\beta$ to another
node is quantified by the thermal Green's function of the network
\citep{estrada2012physics}. This function is better known in the
literature as the communicability function of a graph \citep{estrada2008communicability,estrada2012physics}.
It has found many applications in the analysis of real-world networks,
such as in detecting changes in the contralesional hemisphere following
strokes in humans \citep{crofts2011network}, in the detection of
symptoms of multiple sclerosis \citep{li2013diffusion}, in the study
of variants of epilepsy \citep{campbell2013fusion}, in prediction
of abnormal brain states \citep{iturria2013anatomical}, in early
detection of Alzheimer\textquoteright s disease \citep{mancini2016network},
in prediction of functional protein complexes \citep{ma2012predicting},
in the analysis of genetic diseases \citep{campbell2014netcomm},
in the optimization of wireless networks \citep{chan2016optimizing},
in the evolution of granular materials \citep{walker2010topological},
in the classification of grass pollen \citep{mander2013classification}
and vegetation patterns \citep{mander2017morphometric}, and in the
identification of the transcription factor critically involved with
self-renewal of undifferentiated embryonic stem cells \citep{macarthur2012nanog},
to mention just a few of recent findings.

Melting\textendash the phase transition in which a solid is transformed
into a liquid\textendash is a fundamental physical process of elements,
substances and materials, which results from the application of heat
or pressure to the substance \citep{cahn1986melting,alexiades1992mathematical}.
One of the most successful criteria for explaining melting at the
microscopic level was developed by Lindemann in 1910 \citep{lindemann1910calculation}.
According to Lindemann criterion \citep{lindemann1910calculation,jin2001melting},
melting is caused by vibration instability in the crystal lattice,
which eventually makes that the amplitude of vibration becomes so
large that the atoms collide with their nearest neighbors, disturbing
them and initiating the melting. Then, every substance is characterized
by a melting point, which is the temperature at which such process
starts. A crystal can be represented by a regular lattice \citep{phillpot1989crystals}
in which atoms are the nodes and interactions between atoms are the
edges of a simple graph. It is then easy to set up a vibrational model
on this graph by considering it as a ball-and-spring system and studying
the change of state in it as a result of raising the temperature using
the Lindemann criterion \citep{lindemann1910calculation}. Many granular
materials are nowadays studied by using graph-theoretic methods \citep{papadopoulos2017network}.
Thus, such approach to study melting using graphs is of great importance
in this area of research. However, the most important question that
immediately emerges here is whether we can generalize such theoretical
framework to consider any simple graph. That is, can we use the physical
metaphor of ``melting'' for general graphs and networks? 

In this work we consider a Lindemann-like model for the melting of
graphs and networks. That is, we consider a vibrational model of nodes
in a network based on the communicability function. Then, we prove
analytically the existence of a universal phase transition in the
communicability structure of every simple graph, which resembles the
melting process occurring in substances. We discovered that, similar
to crystalline and amorphous solids, regular and regular-like graphs
``melt'' at lower temperatures and display a sharper transition
between connected to disconnected structures than the random spatial
graphs, which resemble amorphous solids. Finally, we study computationally
this transition in some real-world networks where we observe that
the rate of melting of graphs depends on the topology of the corresponding
network. In particular we observe that this rate changes either as
an exponential or as a power-law with $\beta$. We also discover that
the main driver for node melting is the eigenvector centrality of
the corresponding node. That is, nodes with higher values of the Perron-Frobenious
eigenvector melt at lower temperatures than those with smaller values
of it.

\section{Preliminaries}

\begin{figure}
\centering
\includegraphics[width=0.6\columnwidth,angle=90,clip=0]{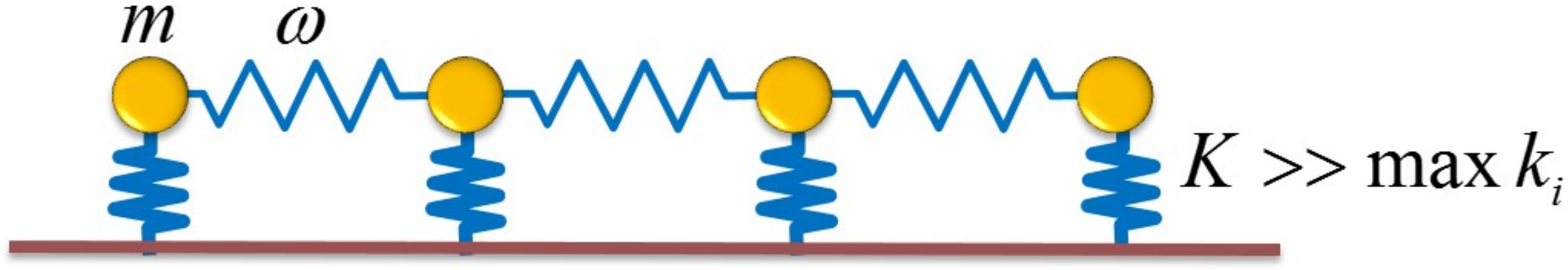}
\caption{Illustration of the model used for deriving the thermal Green's function
of a network of quantum harmonic oscillators.}

\label{vibrational model}
\end{figure}

Here we shall present some definitions, notations, and properties
associated with networks to make this work self-contained. We will
use indistinctly the terms networks and graphs across the paper. Here
we consider only simple, undirected graphs $\Gamma=(V,E)$ with $n$
nodes (vertices) and $m$ edges. The notation used in the paper is
the standard in network theory and the reader is referred to the monograph
\citep{estrada2012structure} for details. An important concept to
be used across this paper is the one of walks. A \textit{walk} of
length $k$ in $\Gamma$ is a set of nodes $i_{1},i_{2},\ldots,i_{k},i_{k+1}$
such that for all $1\leq l\leq k$, $(i_{l},i_{l+1})\in E$. A \textit{closed
walk} is a walk for which $i_{1}=i_{k+1}$. A \textit{path} is a walk
with no repeated nodes. A graph is \textit{connected} if there is
a path connecting every pair of nodes. Let $A$ be the adjacency matrix
of the graph $\Gamma$. For simple graphs $A$ is symmetric and thus
its eigenvalues are real, which we label here in non-increasing order:
$\lambda_{1}\geq\lambda_{2}\geq\ldots\geq\lambda_{n}$. We will consider
the spectral decomposition of $A$: $=U\Lambda U^{T},$ where $\Lambda$
is a diagonal matrix containing the eigenvalues of $A$ and $U=[\mathbf{\overrightarrow{\psi}}_{1},\ldots,\mathbf{\overrightarrow{\psi}}_{n}]$
is orthogonal, where $\mathbf{\overrightarrow{\psi}}_{i}$ is an eigenvector
associated with $\lambda_{i}$. We consider here sets of orthonormalized
eigenvectors of the adjacency matrix. Because the graphs considered
here are connected, $A$ is irreducible and from the Perron-Frobenius
theorem we can deduce that $\lambda_{1}>\lambda_{2}$ and that the
leading eigenvector $\mathbf{\overrightarrow{\psi}}_{1}$ can be chosen
such that its components $\mathbf{\mathbf{\psi}}_{1}(p)$ are positive
for all $p\in V$. It is known that $\left(A^{k}\right)_{pq}$ counts
the number of walks of length $k$ between $p$ and $q$. The following
result concerning the eigenvalue $\lambda_{2}$ is well-known in spectral
graph theory. 
\begin{lem}
\label{lem:Smith}(\citep{smith1970some}) Let $\varGamma=\left(V,E\right)$
be a connected graph and let $\lambda_{1}>\lambda_{2}\geq\lambda_{3}\geq...\geq\lambda_{n}$
be the eigenvalues of $A$. If the graph is not complete multipartite
then $\lambda_{2}>0$. 
\end{lem}
Two important results that we will use in the current work are the
following. For the sake simplicity let us suppose that the vertices
of the graph $\varGamma$ are labeled as $V=\{1,2,3,\ldots,n\}$.
For a given vector $\psi\in\mathbb{\mathbb{R}^{\mathnormal{n}}},$
let $\text{\ensuremath{\mathcal{P}}}(\psi)=\{i:\,\psi(i)>0\}$, $\text{\ensuremath{\mathcal{N}}}(\psi)=\{i:\,\psi(i)<0\}$,
and $\mathcal{O}(\psi)=\{i:\,\psi(i)=0\}$, where $i\in V$, and let
us denote by $\text{\ensuremath{\left\langle \mathcal{P}\left(\psi\right)\right\rangle }, \ensuremath{\left\langle \mathcal{N}\left(\psi\right)\right\rangle }}$
and $\left\langle \mathcal{O}\left(\psi\right)\right\rangle $ the
subgraphs of $\varGamma$ obtained by the nodes of the sets $\text{\ensuremath{\mathcal{P}}}(\psi)$,
$\text{\ensuremath{\mathcal{N}}}(\psi)$ and $\mathcal{O}(\psi)$
respectively.
\begin{lem}
\label{lem:(Theorem-C,)}(\citep{POWERS1988121}) Let $\varGamma=\left(V,E\right)$
be a connected graph. Let $A$ be its adjacency matrix, and let $\lambda_{1}>\lambda_{2}\geq\lambda_{3}\geq...\geq\lambda_{n}$
be the eigenvalues of $A$. Let $(r-1)$ be the multiplicity of $\lambda_{2}$
and let $\psi_{2},\psi_{3},...,\psi_{r}$ be its corresponding eigenvectors.
Suppose that $\cap_{j=2}^{r}\mathcal{O}(\psi_{j})\neq\emptyset$.
Then one of these two cases holds:
\end{lem}
\begin{enumerate}
\item No edge joins a vertex of $\text{\ensuremath{\mathcal{P}}}\left(\psi_{j}\right)$
to one of $\text{\ensuremath{\mathcal{N}\left(\psi_{\mathnormal{j}}\right)}}$,
and $\left\langle \text{\ensuremath{\mathcal{P}}}\left(\psi_{j}\right)\cup\text{\ensuremath{\mathcal{\text{\ensuremath{\mathcal{N}\left(\psi_{\mathnormal{j}}\right)}}}}}\right\rangle $
has $r$ connected components.
\item Some edge joins a vertex of $\text{\ensuremath{\mathcal{P}}}\left(\psi_{j}\right)$
to one of $\text{\ensuremath{\mathcal{N}\left(\psi_{\mathnormal{j}}\right)}}$,
and $\left\langle \text{\ensuremath{\mathcal{P}}}\left(\psi_{j}\right)\cup\text{\ensuremath{\mathcal{\text{\ensuremath{\mathcal{N}\left(\psi_{\mathnormal{j}}\right)}}}}}\right\rangle $,
$\text{\ensuremath{\left\langle \text{\ensuremath{\mathcal{P}}}\left(\psi_{j}\right)\right\rangle }}$
and $\text{\ensuremath{\left\langle \mathcal{\text{\ensuremath{\mathcal{N}\left(\psi_{\mathnormal{j}}\right)}}}\right\rangle }}$
are all connected. 
\end{enumerate}
\begin{lem}
\label{lem:(Theorem-4.2)}(\citep{URSCHEL20141}) Let $\varGamma=\left(V,E\right)$
be a connected graph. Let $\lambda_{2}>0$ be the second largest eigenvalue
of the adjacency matrix with multiplicity $(r-1)$. Then, there exist
eigenvectors $\psi_{2},\psi_{3},...,\psi_{r}$ corresponding to $\lambda_{2}$
such that the induced subgraphs generated by $\text{\ensuremath{\mathcal{P}}}\left(\psi_{j}\right)\cup\mathcal{O}\left(\psi_{j}\right)$
and $\text{\ensuremath{\mathcal{N}}}\left(\psi_{j}\right)$ $\forall j\in\{2,3,\ldots,r\}$
are connected.
\end{lem}
\begin{rem}
\textcolor{black}{Let $\psi$ is an eigenvector corresponding eigenvalue
$\lambda_{2}$. Then $\alpha\psi$, $\alpha\in\mathbb{R}$ is an eigenvector
corresponding $\lambda_{2}$. If $\alpha>0$, then $\text{\ensuremath{\mathcal{P}}}(\alpha\psi)=\text{\ensuremath{\mathcal{P}}}(\psi)$,
$\text{\ensuremath{\mathcal{N}}}(\alpha\psi)=\text{\ensuremath{\mathcal{N}}}(\psi)$
and $\text{\ensuremath{\mathcal{O}}}(\alpha\psi)=\text{\ensuremath{\mathcal{O}}}(\psi)$.
If $\alpha<0$, then $\text{\ensuremath{\mathcal{P}}}(\alpha\psi)=\text{\ensuremath{\mathcal{N}}}(\psi)$,
$\text{\ensuremath{\mathcal{N}}}(\alpha\psi)=\text{\ensuremath{\mathcal{P}}}(\psi)$
and $\text{\ensuremath{\mathcal{O}}}(\alpha\psi)=\text{\ensuremath{\mathcal{O}}}(\psi)$. }
\end{rem}
An important quantity for studying communication processes in networks
has been defined as the communicability function \citep{estrada2008communicability,estrada2010network}. 
\begin{defn}
Let $p$ and $q$ be two nodes of $\Gamma$. The communicability function
between these two nodes is defined as

\[
G_{pq}=\sum_{k=0}^{\infty}\frac{\left(A^{k}\right)_{pq}}{k!}=\left(\exp\left(A\right)\right)_{pq}=\sum_{j=1}^{n}e^{\lambda_{k}}\mathbf{\mathbf{\psi}}_{j}(p)\mathbf{\psi}_{j}(q).
\]
\end{defn}
It counts the total number of walks starting at node $p$ and ending
at node $q$, weighted in decreasing order of their length by a factor
$\frac{1}{k!}$; therefore it is considering shorter walks more influential
than longer ones. In this work we consider a generalization of the
communicability function \citep{estrada2012physics,estrada2007statistical}
consisting of

\[
G_{pq}\left(\beta\right)=\left(\exp\left(\beta A\right)\right)_{pq}=\sum_{j=1}^{n}e^{\beta\lambda_{k}}\mathbf{\mathbf{\psi}}_{j}(p)\mathbf{\psi}_{j}(q),
\]
where $\beta\geq0$ is a parameter that weights homogeneously every
edge of the graph $\Gamma$. Let us give a complete physical interpretation
of the communicability function by considering the following model.
Let us consider a network of quantum-harmonic oscillators, such as
every node is a ball of mass $m$ and two nodes are connected by a
spring of strength constant $\omega$ (see \citep{estrada2012physics}
for details). We tie the network to the ground (to avoid translational
movement) with springs of constant $K\gg\max k_{i}$ as illustrated
in Figure \ref{vibrational model} (we remind that $k_{i}$ is the
degree of the node $i$). The Hamiltonian describing the energy of
this system is given by

\begin{equation}
\hat{H}=\sum_{i}\hbar\varOmega\left(a_{i}^{\dagger}a_{i}+\dfrac{1}{2}\right)-\dfrac{\hbar\omega^{2}}{4\Omega}\sum_{i,j}\left(a_{i}^{\dagger}+a_{i}\right)A_{ij}\left(a_{j}^{\dagger}+a_{j}\right),\label{eq:Hamiltonian}
\end{equation}
where $A_{ij}$ are the elements of the adjacency matrix, $a_{i}^{\dagger}$
($a_{i}$) are the annihilation (creation) operators, and $\Omega=\sqrt{K/m\Omega}$. 

Let us submerge the network of quantum harmonic oscillators into a
thermal bath with inverse temperature $\beta=\left(k_{B}T\right)^{-1},$
where $k_{B}$ is a constant and $T$ is the temperature. Then, the
following result has been previously proved \citep{estrada2012physics}.
\begin{thm}
\textup{\citep{estrada2012physics}} The thermal Green's function
of the network of quantum harmonic oscillators described by \ref{eq:Hamiltonian}
is

\begin{equation}
G_{pq}\left(\beta\right)=\exp\left(-\beta\hbar\Omega\right)\left(\exp\dfrac{\beta\hbar\omega^{2}}{2\Omega}A\right)_{pq}.
\end{equation}
\end{thm}
\begin{rem}
The thermal Green\textquoteright s function accounts for how the node
$p$ (respectively $q$) \textquoteleft feels\textquoteright{} a perturbation
at node $q$ (resp. $p$) due to thermal fluctuations in the bath.
\end{rem}

When the temperature goes to infinity, the inverse temperature $\beta\rightarrow0$,
which means that every edge in the graph vanishes and the resulting
graph is trivial, similar to a system of free particles. When the
temperature goes to zero, the inverse temperature $\beta\rightarrow\infty$
indicating that an infinity number of edges are created between every
pair of nodes connected in $\varGamma$; a situation analogous to
a rigid solid. 

\section{Melting phase transition}

Let us consider a connected graph with eigenvalues ordered as $\lambda_{1}>\lambda_{2}\geq\lambda_{3}\geq...\geq\lambda_{n}$.
Let us then write the communicability function in the following way

\begin{eqnarray*}
G_{pq}(\beta) & = & e^{\beta\lambda_{1}}\psi_{1}\left(p\right)\psi_{1}\left(q\right)+\left[\sum_{2\leq j\leq n}e^{\beta\lambda_{j}}\psi_{j}^{+}\left(p\right)\psi_{j}^{+}\left(q\right)+\sum_{2\leq j\leq n}e^{\beta\lambda_{j}}\psi_{j}^{-}\left(p\right)\psi_{j}^{-}\left(q\right)\right]\\
 &  & +\left[\sum_{2\leq j\leq n}e^{\beta\lambda_{j}}\psi_{j}^{+}\left(p\right)\psi_{j}^{-}\left(q\right)+e^{\beta\lambda_{j}}\psi_{j}^{-}\left(p\right)\psi_{j}^{+}\left(q\right)\right],
\end{eqnarray*}
where $\psi_{j}^{+}\left(p\right)$ ($\psi_{j}^{-}\left(p\right)$)
means that the $p$th entry of the $j$th eigenvector is positive
(negative). Let us consider the graph illustrated in Figure \ref{Vibrations W(2,2)},
where we show the corresponding eigenvectors of the adjacency matrix
in a schematic way. The negative entries of the corresponding eigenvectors
are illustrated like ``vibrations'' in the negative direction of
the $y$- axis. Similarly for the positive entries, which are represented
as vibrations in the positive direction of the $y$- axis. The magnitude
of the vibrations are not represented for the sake of simplicity.
The term $e^{\beta\lambda_{1}}\psi_{1}\left(p\right)\psi_{1}\left(q\right)$
represents the coordinated vibration of all nodes in the graph at
the corresponding value of $\beta$ (see Fig. \ref{Vibrations W(2,2)}),
i.e., a translational motion of the whole network. Then, we obtain
the purely vibrational term for the pairs of nodes as: $\varDelta G_{pq}\left(\beta\right)=G_{pq}\left(\beta\right)-e^{\beta\lambda_{1}}\psi_{1}\left(p\right)\psi_{1}\left(q\right)$,
which can also be expressed as

\begin{equation}
\varDelta G_{pq}(\beta)=\sum_{j\geq2}^{in-phase}e^{\beta\lambda_{j}}\psi_{j}\left(p\right)\psi_{j}\left(q\right)-\left|\sum_{j\geq2}^{out-of-phase}e^{\beta\lambda_{j}}\psi_{j}\left(p\right)\psi_{j}\left(q\right)\right|,
\end{equation}
where the first term, which can be written as $\sum_{2\leq j\leq n}e^{\beta\lambda_{j}}\psi_{j}^{+}\left(p\right)\psi_{j}^{+}\left(q\right)+\sum_{2\leq j\leq n}e^{\beta\lambda_{j}}\psi_{j}^{-}\left(p\right)\psi_{j}^{-}\left(q\right)$,
corresponds to the case when both nodes have the same sign in the
corresponding eigenvector, and the second term, which can be written
as $\sum_{2\leq j\leq n}e^{\beta\lambda_{j}}\psi_{j}^{+}\left(p\right)\psi_{j}^{-}\left(q\right)+e^{\beta\lambda_{j}}\psi_{j}^{-}\left(p\right)\psi_{j}^{+}\left(q\right)$,
accounts for the cases in which the two nodes have different sign
in the corresponding eigenvector. We notice that the second term is
always negative and we use the modulus of it to express the term $\varDelta G_{pq}(\beta)$
as a difference. $\varDelta G_{pq}(\beta)$ accounts for the difference
between the in- and out-of-phase vibrations of the corresponding pair
of nodes. 
\begin{figure}
\centering
\includegraphics[width=0.7\textwidth,angle=90]{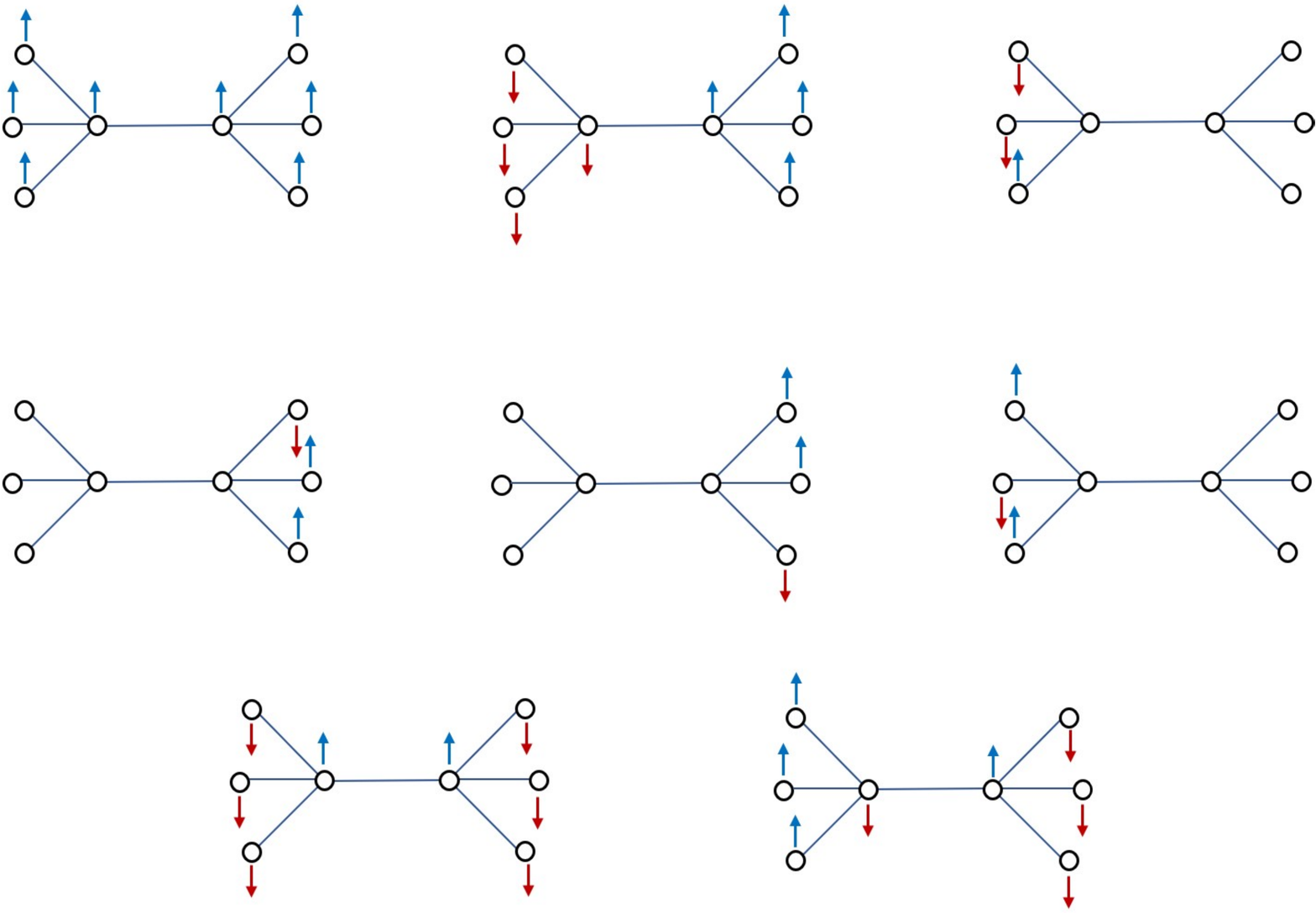}
\caption{Illustration of the sign pattern of the eigenvectors in a simple graph.
Only the signs of the eigenvector components are represented by blue
(positive) and red (negative) arrows. The magnitudes of the eigenvector
components are not represented.}

\label{Vibrations W(2,2)}
\end{figure}

\begin{figure}
\begin{centering}
\includegraphics[width=0.8\textwidth]{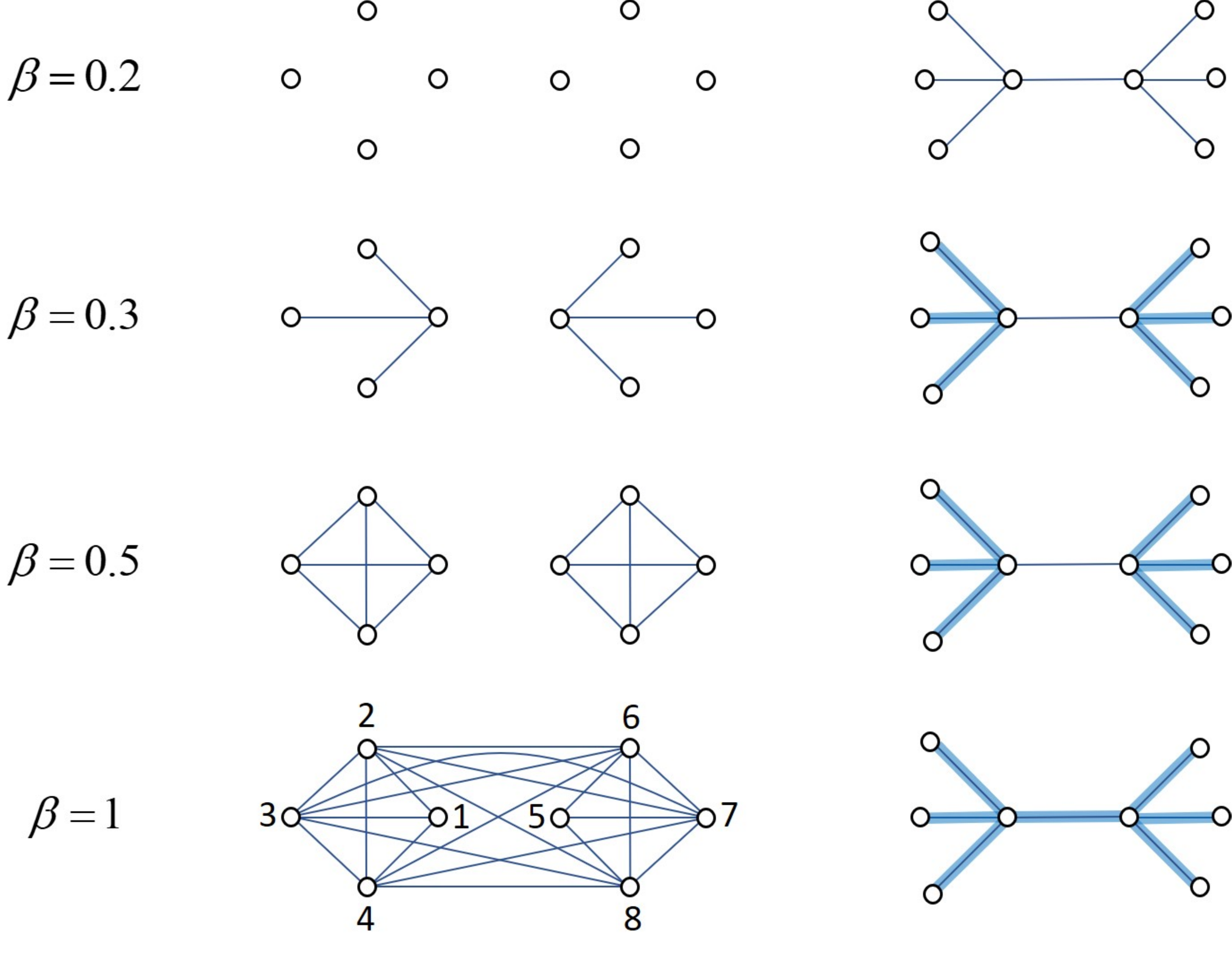}
\par\end{centering}
\caption{Communicability graphs (central panels) at different values of $\beta$
for the tree $\Gamma$ with degree sequence 4,4,1,1,1,1,1,1. In the
right panel the edges of the Lindemann graphs are represented as solid
lines over the edges of the original graph $\Gamma$.}

\label{flow}
\end{figure}

Let us now reconnect with Lindemanncriterion of melting \citep{lindemann1910calculation}
. According to Lindemann the average amplitude of thermal vibrations
in crystals increases with the temperature up to a point in which
the amplitude of vibration is so large that the atoms invade the space
of their nearest neighbors and the melting starts. Lindemann criterion
consists in considering that melting might be expected when the mean-square
amplitude of vibrations exceed a certain threshold value \citep{lindemann1910calculation}.
Let us consider that in a graph $\Gamma$ such threshold is given
by $M(\varGamma,\beta)=\max_{s\neq t\in V}\sum_{j=2}^{n}\psi_{j}\left(s\right)\psi_{j}\left(t\right)e^{\beta\lambda_{j}}$.
That is, that melting starts in a given graph when the vibrations
of the nodes $p$ and $q$ at a given temperature measured by $\Delta G_{pq}\left(\beta\right)$
exceed the value of the maximum vibration of any pair of nodes in
that graph at the same temperature, $M(\varGamma,\beta)$. We should
notice that at a given temperature the terms $\varDelta G_{pq}(\beta)$
and $M(\varGamma,\beta)$ may be either positive or negative. Thus,
in order to implement the Lindemann criterion on graphs we should
sum both terms instead of having their difference,

\begin{equation}
\varDelta\tilde{G}_{pq}(\beta)=M(\varGamma,\beta)+\varDelta G_{pq}(\beta).
\end{equation}
Then, when $M(\varGamma,\beta)>0$ we have the following scenarios.
If $\varDelta G_{pq}(\beta)>0$ then $\varDelta\tilde{G}_{pq}(\beta)$
is always positive as it is the sum of two positive terms indicating
a reinforcement of the in-phase vibrations of the two nodes. If $\varDelta G_{pq}(\beta)<0$
then $\varDelta\tilde{G}_{pq}(\beta)>0$ if the difference between
the in-phase and out-of-phase vibrations $\varDelta G_{pq}(\beta)$,
does not overtake the maximum in-phase vibrations of any pair of nodes
in the graph. Otherwise, $\varDelta\tilde{G}_{pq}(\beta)<0$, which
indicates that the out-of-phase vibrations of these two nodes have
overtaken not only their in-phase vibrations but also the maximum
in-phase vibrations of any pair of nodes in the graph. In this last
case we will say that the corresponding edge has been melted. On the
other hand, when $M(\varGamma,\beta)<0$, then also $\varDelta G_{pq}(\beta)<0$
which means that $\varDelta\tilde{G}_{pq}(\beta)<0$ and the edge
necessarily melts. Then, we will call $\varDelta\tilde{G}_{pq}(\beta)$
the \textit{graph Lindemann criterion}, having in mind that the melting
will start when $\varDelta\tilde{G}_{pq}(\beta)<0$. Let us define
the following representation of $\varDelta\tilde{G}_{pq}(\beta)$
in the form of a new graph.
\begin{defn}
Let $\Gamma=\left(V,E\right)$ be a simple graph. The \textit{communicability
graph} $H\left(V,E',\beta\right)$ of $\Gamma=\left(V,E\right)$ is
the graph with the same set of nodes as $\Gamma$ and with edge set
given by the following adjacency relation 

\begin{equation}
A(H,\beta)_{p,q}=\left\{ \begin{array}{c}
\textnormal{1 if }\varDelta\tilde{G}_{pq}(\beta)\geq0,\\
\textnormal{0 if }\varDelta\tilde{G}_{pq}(\beta)<0.
\end{array}\right.
\end{equation}
\end{defn}
In the communicability graph there could be edges connecting pairs
of nodes which are not connected in the original graph $\Gamma.$
In a similar way, there could be pairs of nodes not connected in $H\left(V,E',\beta\right)$
which correspond to edges in $\Gamma$ (see further example). In other
words, $\Gamma$ is not necessarily a subgraph of $H\left(V,E',\beta\right)$.
For instance, in Fig. \ref{flow} at $\beta=1$ the two central nodes
1 and 5 of the graph are vibrating out-of-phase. However, because
we do not have a temporal sequence of how the vibrations occurs there
are also paths connecting 1 and 5 in which all the nodes vibrate in
phase. This is the case of the paths 1-2-6-5, 1-4-8-5 and so forth.
As a consequence of these paths the nodes 1 and 5 can be vibrating
in-phase at some temporal stages of the process. For that reason we
introduce the following definitions.
\begin{defn}
Let $\Gamma=\left(V,E\right)$ be a simple graph and let $H\left(V,E',\beta\right)$
be its communicability graph. Let $p$ and $q$ be two nodes of $\Gamma.$
We say that there is a \textit{Lindemann path }between the nodes $p$
and $q$ in $\Gamma$ at a given value of $\beta$ if there is a path
connecting both nodes in the communicability graph $H\left(V,E',\beta\right)$.
In this case we say that $\exists L_{p,q}$. Otherwise, we say that
$\nexists L_{p,q}$. 
\end{defn}
We now define a graph that contain all the information about the in-
and out-phase nature of the vibrations in a graph $\Gamma=\left(V,E\right)$. 
\begin{defn}
Let $\Gamma=\left(V,E\right)$ be a simple graph and let $H\left(V,E',\beta\right)$
be its communicability graph. The \textit{Lindemann graph} $F\left(V,E'',\beta\right)$
of $\Gamma$ is the graph with the same set of vertices as $\Gamma$
and edge set defined by the following adjacency relation

\begin{equation}
A(F,\beta)_{p,q}=\left\{ \begin{array}{c}
\textnormal{1 if \ensuremath{\left(p,q\right)\in E} and \ensuremath{\exists L_{p,q}}},\\
\textnormal{0 if \ensuremath{\left(p,q\right)\notin E} or \ensuremath{\nexists L_{p,q}}}.
\end{array}\right.
\end{equation}
\end{defn}
To illustrate the previously defined concepts we return to the tree
with eight nodes and degree sequence 4,4,1,1,1,1,1,1 at different
values of $\beta$ illustrated in Fig. \ref{flow}. For $\beta=1$
the communicability graph has many more edges than the original tree
$\Gamma,$ but it also misses the central link connecting the nodes
1 and 5. When constructing the Lindemann graph we should observe that
there is a path between every pair of nodes in the corresponding communicability
graph, i.e., it is connected. Thus, the Lindemann graph consists of
the same set of edges as the original graph. The Lindemann graph is
represented by solid lines in the right panels of Fig. \ref{flow}.
When $\beta=0.5$ the communicability graph consists of two cliques
of four nodes each. Then, the Lindemann graph consists of all edges
of $\Gamma$, except the central edge connecting the nodes 1 and 5,
because there is no path connecting the two nodes of degree 4 in the
communicability graph. At this value of $\beta$ we can say that the
melting of this graph has started because for (at least) one pair
of nodes the out-of-phase vibrations have overcome the graph Lindemann
criterion. Notice that for $\beta=0.3$ the communicability graph
has changed in respect to that for $\beta=0.5$, but the Lindemann
graphs are exactly the same due to the double conditions that need
to be required for having an edge in these graphs. Finally, when $\beta=0.2$
the communicability graph is formed by 8 isolated nodes and so is
the Lindemann graph. At this point there is no communicability between
any pair of nodes and the Lindemann graph is the trivial graph. In
our physical metaphor, the graph is totally ``melted''.

\begin{figure}
\begin{centering}
\includegraphics[width=1\textwidth]{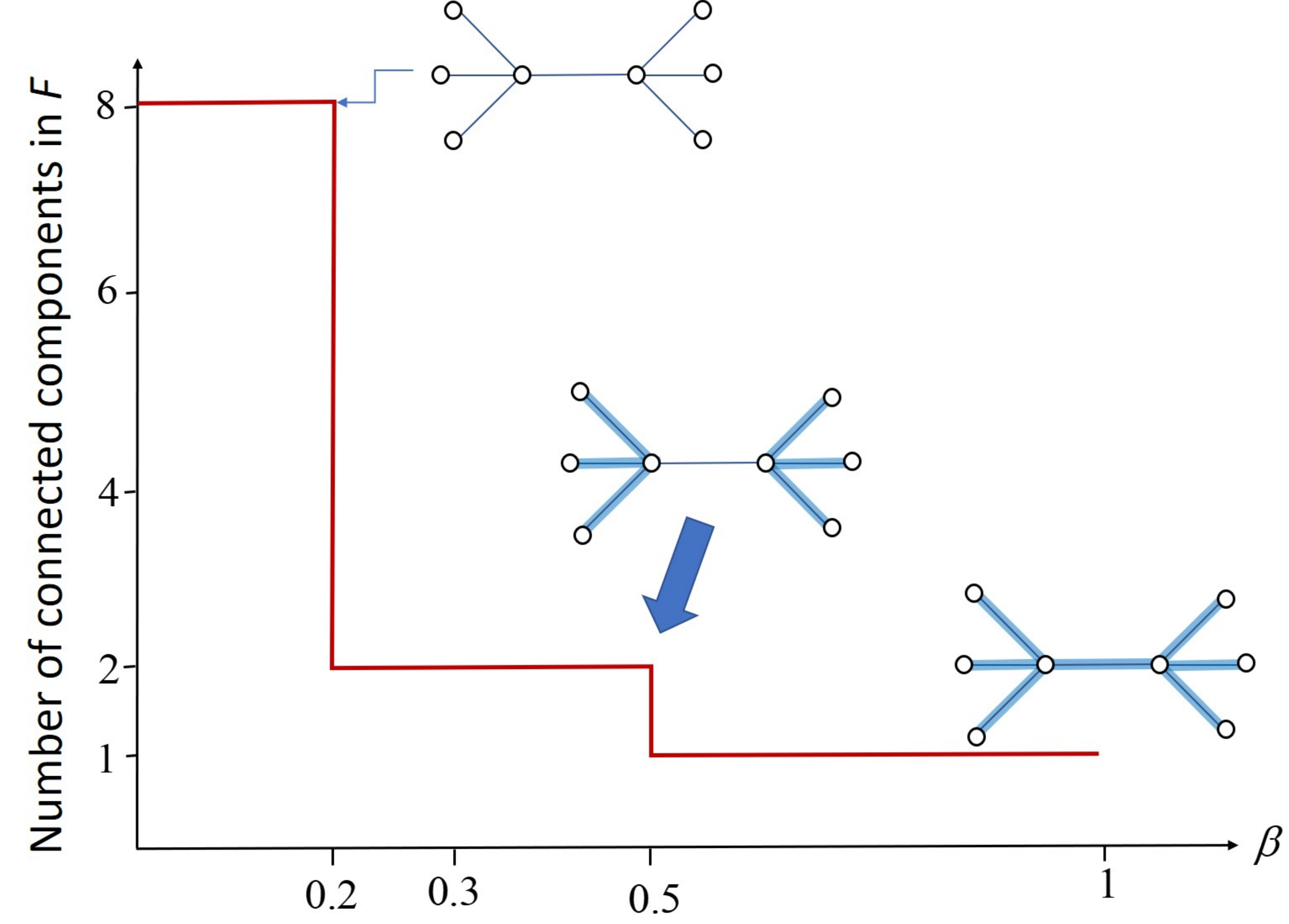}
\par\end{centering}
\caption{Illustration of the transition between connected $\beta>0.5$ to disconnected
$\beta\leq0.5$ Lindemann graph as a function of $\beta$ for the
simple tree illustrated in Fig. \ref{flow}. }

\label{sketch}
\end{figure}

In the Fig. \ref{sketch} we have plotted the values of $\beta$ versus
the number of connected components of the Lindemann graph. At the
point $\beta=0.5$, marked in the plot with a fat arrow, there is
a transition between a connected to a disconnected Lindemann graph.
We have previously identified this value of $\beta=\beta_{c}$ as
the melting temperature of this graph. The reader should keep in mind
that the physical terms used here represent metaphors from the physical
world to a mathematical framework and not a physical reality. The
question that immediately emerges here is whether this phase transition
is universal for any simple graph or not. In the next section we respond
positively to this question.

\section{Main result}

Here we prove the existence of the phase transition between connected
to disconnected Lindemann graph for any simple graph. For proving
this result it is enough to prove that this transition occurs in the
communicability graph. Then, when the communicability graph is connected
also is the Lindemann graph, due to the fact that there is a path
between every pair of nodes in a connected graph. In the same way,
if the communicability graph is disconnected also is the Lindemann
graph because there will be pairs of adjacent nodes of the original
graph for which there are no paths connecting them in the communicability
graph. We divide our result into two parts. The first deals with all
graphs which are not complete multipartite ones. The second proves
the result for this kind of graphs.
\begin{thm}
\label{thm:main}Let $H(\varGamma,\beta)$ be the communicability
graph for a non complete multipartite graph $\Gamma=\left(V,E\right).$
Then, there exist a value $\beta_{c}\in[0,\infty)$ such that

(i) \textup{$H\left(\varGamma,\beta\geq\beta_{c}\right)$ is connected;}

(ii) \textup{$H\left(\varGamma,\beta<\beta_{c}\right)$ is disconnected.}\\
\begin{proof}
We start by proving that the communicability graph $H(\varGamma,\beta)$
is disconnected for certain value of the inverse temperature. Then,
we prove that it becomes connected for certain value of $\beta$,
which immediately implies that the communicability graph $H(\varGamma,\beta)$
makes a transition from connected to disconnected at certain intermediate
temperature, which we call $\beta_{c}$. This value is unique since
the communicability function is monotonic, as it is the sum of exponential
functions which are monotonic. The communicability graph function
is

\begin{equation}
\varDelta\tilde{G}_{pq}(\beta):=M(\varGamma,\beta)+\sum_{j=2}^{n}\psi_{j}\left(p\right)\psi_{j}\left(q\right)e^{\beta\lambda_{j}},
\end{equation}
where $M(\varGamma,\beta)=\max_{s\neq t}\sum\psi_{j}\left(s\right)\psi_{j}\left(t\right)e^{\beta\lambda_{j}}$.
Two distinct nodes $p\neq q$ are connected in $H(\varGamma,\beta)$
if $\triangle\tilde{G}_{pq}\geq0$, and disconnected if $\triangle\tilde{G}_{pq}<0$.
Let us consider the case when $\beta\rightarrow0$. In this case we
have $\lim_{\beta\rightarrow0}\sum_{j=2}^{n}\psi_{j}\left(p\right)\psi_{j}\left(q\right)e^{\beta\lambda_{j}}=\sum_{j=2}^{n}\psi_{j}\left(p\right)\psi_{j}\left(q\right)=-\psi_{1}\left(p\right)\psi_{1}\left(q\right)$.
Thus,

\begin{equation}
\triangle\tilde{G}_{pq}\left(\beta\rightarrow0\right)=-\left(\psi_{1}\left(p\right)\psi_{1}\left(q\right)+\max_{s\neq t}\psi_{1}\left(s\right)\psi_{1}\left(t\right)\right)<0.
\end{equation}
Then, it is obvious that the edge $pq$ in the graph is disconnected
and this happens for every pair of nodes in the graph. Consequently,
the communicability graph is disconnected for some $\beta$ when $\beta\rightarrow0$. 

Let us now consider that case when $\beta\rightarrow\infty$. in this
case the communicability function $\triangle\tilde{G}_{pq}(\beta)$
is dominated by the term containing the second largest eigenvalue
$\lambda_{2}$. Now let $\left(r-1\right)\geq1$ be the multiplicity
of $\lambda_{2}$, i.e., $\lambda_{2}=\lambda_{3}=\cdots=\lambda_{r}$.
Then, $\forall p,q\in V,$ we have 
\begin{eqnarray*}
\triangle\tilde{G}_{pq}(\beta) & \rightarrow & (\psi_{r}\left(p\right)\psi_{r}\left(q\right)+\psi_{r-1}\left(p\right)\psi_{r-1}\left(q\right)+\ldots+\psi_{2}\left(p\right)\psi_{2}\left(q\right))e^{\beta\lambda_{2}}\\
 &  & +\max_{s\neq t}\left(\psi_{r}\left(s\right)\psi_{r}\left(t\right)+\psi_{r-1}\left(s\right)\psi_{r-1}\left(t\right)+\ldots+\psi_{2}\left(s\right)\psi_{2}\left(t\right)\right)e^{\beta\lambda_{2}},
\end{eqnarray*}
\begin{equation}
\Rightarrow\triangle\tilde{G}_{pq}(\beta)\rightarrow e^{\beta\lambda_{2}}\sum_{j=2}^{r}\psi_{j}\left(p\right)\psi_{j}\left(q\right)+e^{\beta\lambda_{2}}\max_{s\neq t}\sum_{j=2}^{r}\psi_{j}\left(s\right)\psi_{j}\left(t\right),\label{eq:1}
\end{equation}
where $\psi_{2},\psi_{3},\ldots,\psi_{r-1},\psi_{r}$, are the eigenvectors
corresponding to the eigenvalue $\lambda_{2}$. \\
If $\cap_{j=2}^{r}\mathcal{O}(\psi_{j})\neq\emptyset$,\textcolor{black}{{}
then by Lemma \ref{lem:(Theorem-C,)}, one of the following two separate
cases hold.}
\begin{casenv}
\item No edge in the original graph $\varGamma=\left(V,E\right)$ joins
a vertex of $\text{\ensuremath{\mathcal{P}}}\left(\psi_{j}\right)$
to one of $\text{\ensuremath{\mathcal{N}\left(\psi_{\mathnormal{j}}\right)}}$.
\textcolor{black}{Then $\psi_{j}\left(p\right)\psi_{j}\left(q\right)\geq0$,
}$\forall j\in\{2,3,\ldots,r\},$ \textcolor{black}{and $\forall(p,q)\in E$,
}$\triangle\tilde{G}_{pq}$\textcolor{black}{{} (\ref{eq:1}) }satisfies:
\\
\begin{equation}
\triangle\tilde{G}_{pq}(\beta)\geq0.\label{eq:2}
\end{equation}
Now let us rewrite the function $\triangle\tilde{G}_{pq}$ in the
following form, where $s$ and $t$ are two distinct nodes in the
graph:
\[
\triangle\tilde{G}_{pq}=\left(\exp\left(\beta A\right)\right){}_{pq}-\psi_{1}\left(p\right)\psi_{1}\left(q\right)e^{\beta\lambda_{1}}+\max_{s\neq t}\left[\left(\exp\left(\beta A\right)\right){}_{st}-\psi_{1}\left(s\right)\psi_{1}\left(t\right)e^{\beta\lambda_{1}}\right].
\]
Let us now consider $\beta=0$, then $\forall p\neq q\in V,$ \\
\begin{equation}
\triangle\tilde{G}_{pq}=-\psi_{1}\left(p\right)\psi_{1}\left(q\right)+\max_{s\neq t}\left[-\psi_{1}\left(s\right)\psi_{1}\left(t\right)\right]<0,\label{eq:3}
\end{equation}
since the leading eigenvector $\vec{\psi}_{1}$ can be chosen such
that its components $\mathbf{\mathbf{\psi}}_{1}(p)$ are positive
for all $p\in V$ according to the Perron-Frobenius theorem. The exponential
matrix $\exp(0\cdot A)=I$, so that $(\exp(\beta A))_{pq}=0$, $\forall p\neq q\in V.$
Then, from \ref{eq:2} and \ref{eq:3}, we have that there is a value
of $\beta_{c}\in[0,\infty),$ such that $H\left(\varGamma,\beta\geq\beta_{c}\right)$
is connected and $H\left(\varGamma,\beta<\beta_{c}\right)$ is disconnected.\\
\item There exists at least one edge that joins a vertex of $\text{\ensuremath{\mathcal{P}}}(\psi_{j})$
to one of $\text{\ensuremath{\mathcal{N}}(\ensuremath{\psi_{j}})}$,
moreover $\left\langle \text{\ensuremath{\mathcal{P}}}(\psi_{j})\right\rangle $
and $\left\langle \text{\ensuremath{\mathcal{N}}}(\psi_{j})\right\rangle $
are connected.\\
\textcolor{black}{{} Let $p\in\cap_{j=2}^{r}\mathcal{O}(\psi_{j})$,}
\textcolor{black}{then $\psi_{j}\left(p\right)=0$, }$\forall j\in\{2,3,\ldots,r\}.$
\textcolor{black}{Then} $\forall q\neq p\in V,$ we get\textcolor{black}{:
\[
e^{\beta\lambda_{2}}\sum_{j=2}^{r}\psi_{j}\left(p\right)\psi_{j}\left(q\right)=0,
\]
and }\\
\begin{equation}
\triangle\tilde{G}_{pq}(\beta)\geq0.\label{eq:5}
\end{equation}
Therefore all the nodes of $V$ are connected to each other through
$p$. Then from \ref{eq:3}, and \ref{eq:5}, we have that there is
$\exists\beta_{c}\in[0,\infty),$ such that $H\left(\varGamma,\beta\geq\beta_{c}\right)$
is connected and $H\left(\varGamma,\beta<\beta_{c}\right)$ is disconnected. 
\end{casenv}
Now let us consider the case in which $\cap_{j=2}^{r}\mathcal{O}(\psi_{j})=\emptyset$.
Let $p\in\text{\ensuremath{\mathcal{P}}(}\psi_{j})\cup\mathcal{O}(\psi_{j})$
(resp., $p\in\text{\text{\ensuremath{\mathcal{N}}}(}\psi_{j}))$ $j\in\{2,3,\ldots,r\}$.\textcolor{black}{{}
Then according to Lemma \ref{lem:(Theorem-4.2)}, there exists }$q\in\text{\ensuremath{\mathcal{P}}(}\psi_{j})\cup\mathcal{O}(\psi_{j})$
(resp., $q\in\text{\text{\ensuremath{\mathcal{N}}}(}\psi_{j}))$\textcolor{black}{{}
such that $(p,q)\in E$ and either }$p,q\in\text{\ensuremath{\mathcal{P}}(}\psi_{j})\cup\mathcal{O}(\psi_{j})$
or $\text{\ensuremath{\mathcal{N}}}(\psi_{j})$, $\forall j\in\{2,\ldots,r\}$.
It holds that\textcolor{black}{{} $\psi_{j}\left(p\right)\psi_{j}\left(q\right)\geq0$,
}$\forall j$ such that\textcolor{black}{{} }we get:\textcolor{black}{
\[
\sum_{j=2}^{r}\psi_{j}\left(p\right)\psi_{j}\left(q\right)\geq0,
\]
and }\\
\begin{equation}
\triangle\tilde{G}_{pq}(\beta)\geq0.\label{eq:6}
\end{equation}
\textcolor{black}{Now, since there are $(r-1)$ eigenvectors corresponding
to the eigenvalue $\lambda_{2}$, there are $2^{r-1}$ of intersected
subsets in $V$ of $\text{\ensuremath{\mathcal{P}}(}\psi_{j})\cup\mathcal{O}(\psi_{j})$
and $\text{\ensuremath{\mathcal{N}}}(\psi_{j})$. In these sets the
nodes have the same sings of the eigenvector components $\psi_{j}$,
$\forall j\in\{2,3,\ldots,r\}.$ Let us denote these subsets by $W_{1},W_{2},\cdots,W_{2^{r-1}}$
( When $r-1=2$ these subsets are: $W_{1}=(\text{\ensuremath{\mathcal{P}}(}\psi_{2})\cup\mathcal{O}(\psi_{2}))\cap(\text{\ensuremath{\mathcal{P}}(}\psi_{3})\cup\mathcal{O}(\psi_{3}))$,
$W_{2}=\text{(\ensuremath{\mathcal{P}}(}\psi_{2})\cup\mathcal{O}(\psi_{2}))\cap\text{\ensuremath{\mathcal{N}}}(\psi_{3})$,
$W_{3}=\text{\ensuremath{\mathcal{N}}}(\psi_{2})\cap\text{(\ensuremath{\mathcal{P}}(}\psi_{3})\cup\mathcal{O}(\psi_{3}))$
and $W_{4}=\text{\ensuremath{\mathcal{N}}}(\psi_{2})\cap\text{\ensuremath{\mathcal{N}}(}\psi_{3})$).
Then $\forall p,q\in W_{h}$, $\forall h\in\{1,2,\ldots,2^{r-1}\}$,
it holds $\psi_{j}\left(p\right)\psi_{j}\left(q\right)\geq0$, $\forall j$
and $\triangle\tilde{G}_{pq}(\beta)\geq0.$ So that the supgraph $\left\langle W_{h}\right\rangle $,
$\forall h\in\{1,2,\ldots,2^{r-1}\}$, is connected.}
\begin{proof}
\textcolor{black}{Finally, we need to show whether the subgraphs }$\left\langle W_{1}\right\rangle ,\left\langle W_{2}\right\rangle ,\cdots,\left\langle W_{2^{r-1}}\right\rangle $
\textcolor{black}{are connected to each other, such that we get a
connected graph.} Let $p'\in\left\langle W_{h}\right\rangle $ and
$q'\in\left\langle W_{s}\right\rangle $, where\textcolor{black}{{}
}$h,s\in\{1,2,\ldots,2^{r-1}\}$,\textcolor{black}{{} }such that the
absolute value of $\sum_{j=2}^{r}\psi_{j}\left(p'\right)\psi_{j}\left(q'\right)$,
satisfies: 
\[
\sum_{j=2}^{r}\psi_{j}\left(p'\right)\psi_{j}\left(q'\right)\leq\max_{s\neq t\in V}\sum_{j=2}^{r}\psi_{j}\left(s\right)\psi_{j}\left(t\right),
\]
\textcolor{black}{Then }$\triangle\tilde{G}_{p'q'}$\textcolor{black}{{}
(\ref{eq:1}) }satisfies: \\
\begin{equation}
\triangle\tilde{G}_{p'q'}(\beta)\geq0,\label{eq:7}
\end{equation}
therefore $\left\langle W_{1}\right\rangle ,\left\langle W_{2}\right\rangle ,\cdots,\left\langle W_{2^{r-1}}\right\rangle $
are connected to each other. Then from \ref{eq:3}, \ref{eq:6} and
\ref{eq:7} we have there is $\beta_{c}\in[0,\infty),$ such that
$H\left(\varGamma,\beta\geq\beta_{c}\right)$ is connected and $H\left(\varGamma,\beta<\beta_{c}\right)$
is disconnected, which finally proves the result.
\end{proof}
\end{proof}
\end{thm}
\begin{rem}
\textcolor{black}{In the case of complete multipartite graphs which
were not included in the Theorem \ref{thm:main} we have the following.
As in the general case $\varDelta\tilde{G}_{pq}(\beta\rightarrow0)<0$,
which indicates that the edge does not exist, and this is true for
any pair of nodes in the graph. Then, we have to show when such edge
exists. According to Lemma \ref{lem:Smith} \citep{smith1970some}
in complete multipartite graphs $\lambda_{2}\leq0$. Then, let us
first consider the case when $\lambda_{2}=0$. }In this case when
$\beta\rightarrow\infty$, the communicability function $\triangle\tilde{G}_{pq}(\beta)$
is dominated by the term of the largest eigenvalue $\lambda_{2}$.
Now let $\left(r-1\right)\geq1$ be the multiplicity of $\lambda_{2}$,
i.e., $\lambda_{2}=\lambda_{3}=\cdots=\lambda_{r}$. Then $\forall p,q\in V,$
\\
\begin{eqnarray*}
\triangle\tilde{G}_{pq}(\beta) & \rightarrow & e^{\beta\lambda_{2}}\sum_{j=2}^{r}\psi_{j}\left(p\right)\psi_{j}\left(q\right)+e^{\beta\lambda_{2}}\max_{s\neq t\in V}\sum_{j=2}^{r}\psi_{j}\left(s\right)\psi_{j}\left(t\right),
\end{eqnarray*}
which implies that
\begin{rem}
\begin{equation}
\triangle\tilde{G}_{pq}(\beta)\rightarrow\sum_{j=2}^{r}\psi_{j}\left(p\right)\psi_{j}\left(q\right)+\max_{s\neq t\in V}\sum_{j=2}^{r}\psi_{j}\left(s\right)\psi_{j}\left(t\right),\label{eq:Lindemann}
\end{equation}
since $e^{\beta\lambda_{2}}=1$. So that the proof for the case when
$\cap_{j=2}^{r}\mathcal{O}(\psi_{j})\neq\emptyset$, will be the same
as that for the case when $\lambda_{2}>0$, which is the Case 2, in
Theorem 10.\\
Now when $\cap_{j=2}^{r}\mathcal{O}(\psi_{j})=\emptyset$ then for
the nodes which belong to the sets $W_{h}$, $\forall h\in\{1,2,\ldots,2^{r-1}\}$,
(the sets of all of intersected subsets in $V$ of $\text{\ensuremath{\mathcal{P}}(}\psi_{j})\cup\mathcal{O}(\psi_{j})$
and $\text{\ensuremath{\mathcal{N}}}(\psi_{j})$, $j\in\{2,3,\ldots,r\}$)
are connected to each other in $\left\langle W_{h}\right\rangle $,
since they have the same sings of the eigenvector components $\psi_{j}$,
$\forall j\in\{2,3,\ldots,r\}$. For the nodes which do not belong
to $W_{h}$, $\forall h\in\{1,2,\ldots,2^{r-1}\}$, let $p\in V$,
$p\notin W_{h}$, $\forall h\in\{1,2,\ldots,2^{r-1}\}$, and let $q\in W_{h}$,
such that the absolute value of $\sum_{j=2}^{r}\psi_{j}\left(p\right)\psi_{j}\left(q\right)$,
satisfies
\[
\left|\sum_{j=2}^{r}\psi_{j}\left(p\right)\psi_{j}\left(q\right)\right|\leq\max_{s\neq t\in V}\sum_{j=2}^{r}\psi_{j}\left(s\right)\psi_{j}\left(t\right),
\]
\textcolor{black}{Then }$\triangle\tilde{G}_{pq}$\textcolor{black}{{}
(\ref{eq:Lindemann}) }satisfies 

\begin{equation}
\triangle\tilde{G}_{pq}(\beta)\geq0.\label{eq:cond_1}
\end{equation}
So let us denote the subgraphs generated by $W_{h}$, and the other
nodes of $V$ which do not belong to $W_{h}$, by\textcolor{black}{{}
}$\left\langle W'_{1}\right\rangle ,\left\langle W'_{2}\right\rangle ,\cdots,\left\langle W'_{2^{r-1}}\right\rangle $.
\textcolor{black}{Finally, in the same previous way we can connect
}$\left\langle W'_{1}\right\rangle ,\left\langle W'_{2}\right\rangle ,\cdots,\left\langle W'_{2^{r-1}}\right\rangle $
\textcolor{black}{to each other, such that we get a connected graph.}
Let $p'\in\left\langle W'_{h}\right\rangle $ and $q'\in\left\langle W'_{s}\right\rangle $,
where\textcolor{black}{{} }$h,s\in\{1,2,\ldots,2^{r-1}\}$,\textcolor{black}{{}
}such that the absolute value of $\sum_{j=2}^{r}\psi_{j}\left(p'\right)\psi_{j}\left(q'\right)$,
satisfies: \\
\[
\left|\sum_{j=2}^{r}\psi_{j}\left(p'\right)\psi_{j}\left(q'\right)\right|\leq\max_{s\neq t\in V}\sum_{j=2}^{r}\psi_{j}\left(s\right)\psi_{j}\left(t\right),
\]
\textcolor{black}{Then }$\triangle\tilde{G}_{pq}$\textcolor{black}{{}
(\ref{eq:Lindemann}) }satisfies 

\begin{equation}
\triangle\tilde{G}_{p'q'}(\beta)\geq0.\label{eq:cond_2}
\end{equation}
Therefore $\left\langle W'_{1}\right\rangle ,\left\langle W'_{2}\right\rangle ,\cdots,\left\langle W'_{2^{r-1}}\right\rangle $
are connected to each other. Then from \ref{eq:3}, \ref{eq:cond_1}
and \ref{eq:cond_2} we have that there is $\beta_{c}\in[0,\infty),$
such that $H\left(\varGamma,\beta\geq\beta_{c}\right)$ is connected
and $H\left(\varGamma,\beta<\beta_{c}\right)$ is disconnected.

\textcolor{black}{On the other hand, when $\lambda_{2}<0$ we have
that }

\textcolor{black}{
\begin{equation}
\varDelta\tilde{G}_{pq}(\beta)=\max_{s\neq t}\sum\psi_{j}\left(s\right)\psi_{j}\left(t\right)e^{-\beta\left|\lambda_{j}\right|}+\sum_{j=2}^{n}\psi_{j}\left(p\right)\psi_{j}\left(q\right)e^{-\beta\left|\lambda_{j}\right|}.
\end{equation}
}This means that $\lim_{\beta\rightarrow\infty}\varDelta\tilde{G}_{pq}(\beta)=0$.
However, such limit can be either positive\textendash existence of
the edge\textendash or negative\textendash not existence of the edge.
Thus, in this case we can consider that the edge exists if $\left|\lim_{\beta\rightarrow\infty}\varDelta\tilde{G}_{pq}(\beta)\right|\leq\varepsilon$,
where $\varepsilon$ is a threshold very close to zero. \textcolor{black}{The
case of complete multipartite graphs is an all-or-nothing case of
melting. That is, at $\beta=0$ all the nodes in the Lindemann graph
are isolated. When $\beta$ is very large all the nodes in the communicability
graph are connected to each other and the Lindemann graph is the same
as the original graph. Thus, in these graphs there is not a gradual
melting as in the rest of the graphs but an abrupt transition between
being connected to fully disconnected occurs.}
\end{rem}
\end{rem}

\section{Modeling results}

\subsection{Melting of granular materials}

\begin{figure}
\centering
\includegraphics[width=0.35\textwidth]{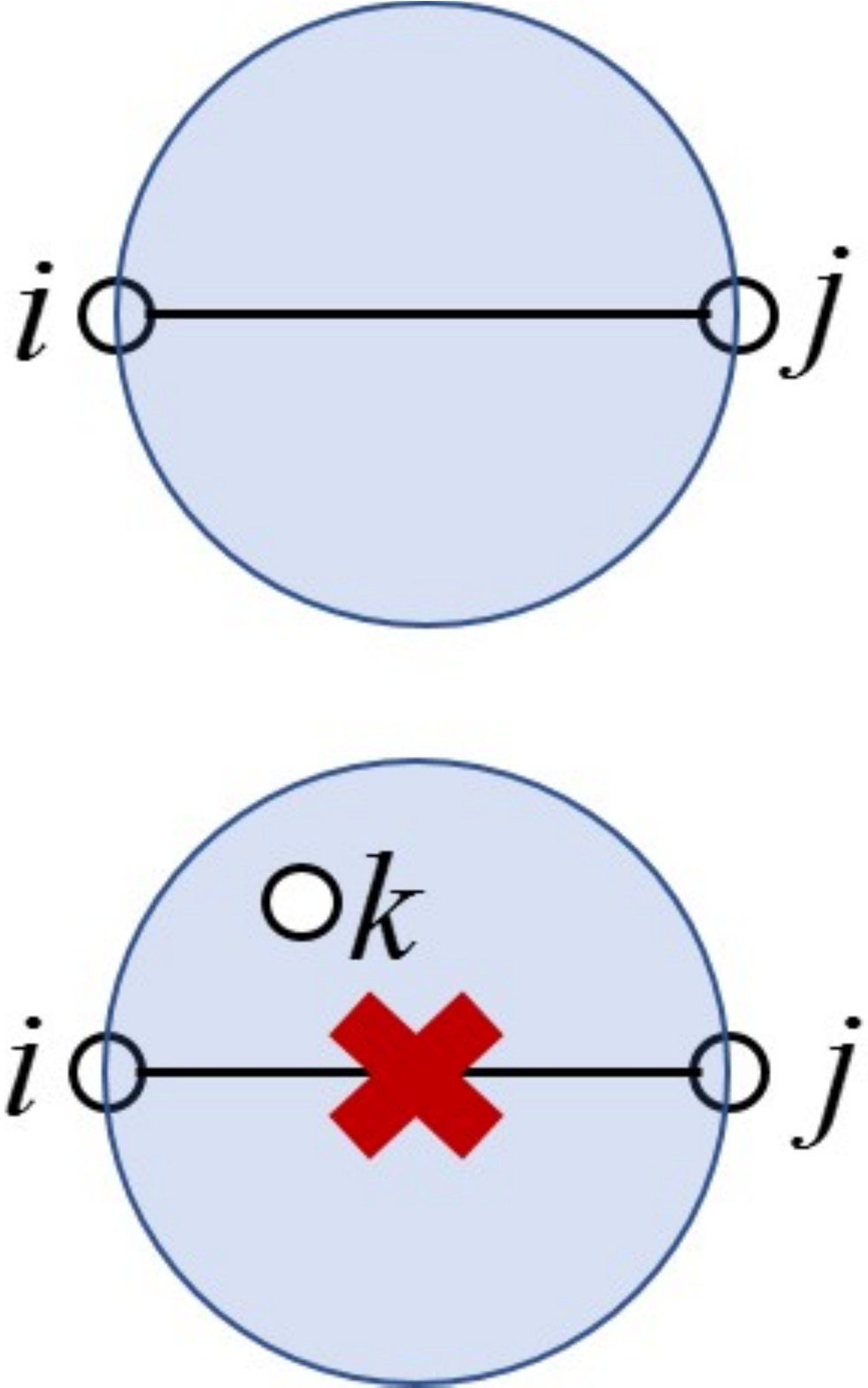}
\includegraphics[width=0.35\textwidth]{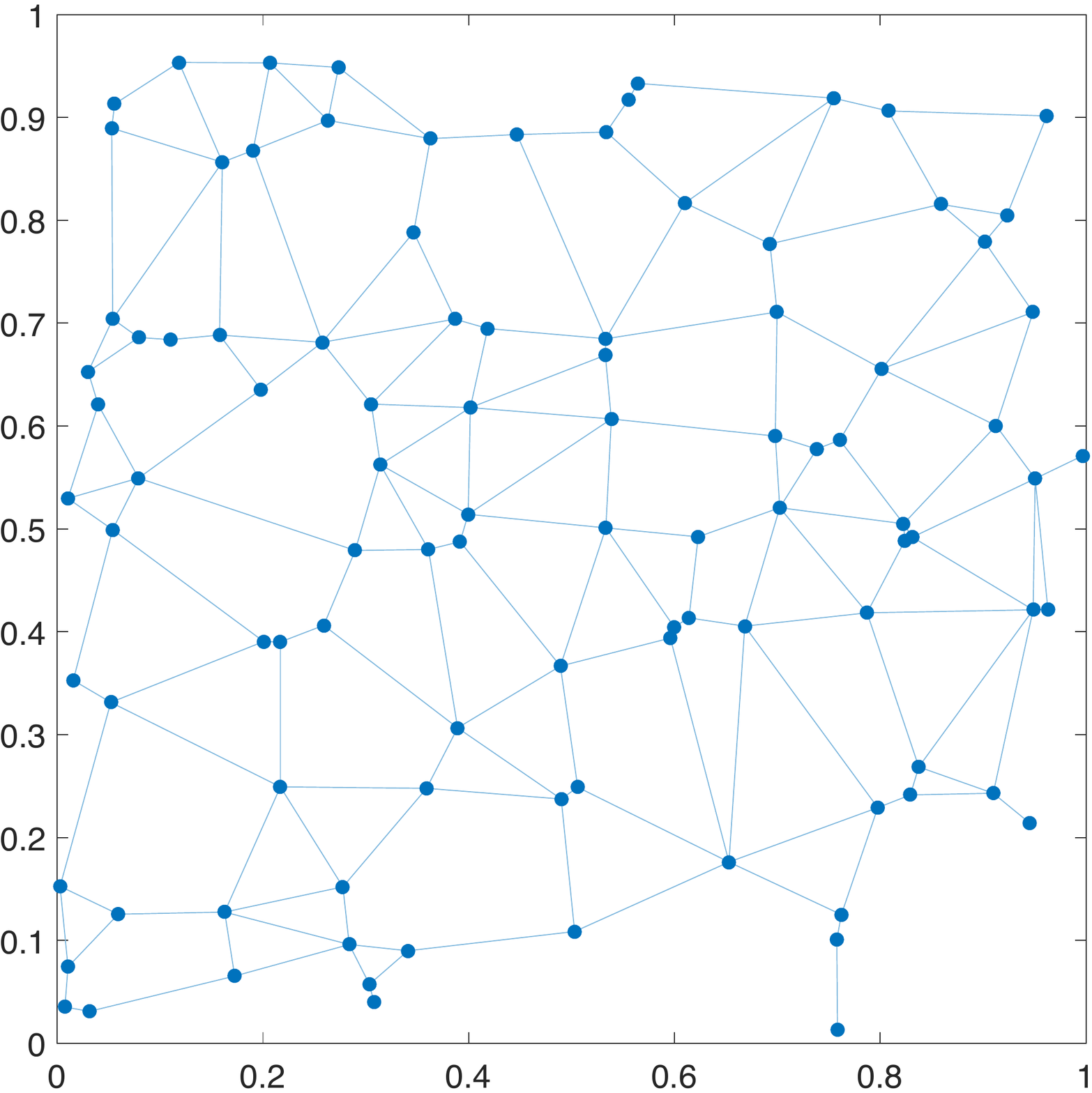}
\caption{Illustration of the construction process of a Gabriel graph (left),
where a disk is defined for a pair of nodes which forms a diameter
of the disk. Because there is no point inside that disk the two nodes
are connected (top graphic). In the bottom graphic a point $k$ is
inside the disk and the two nodes $i$ and $j$ are not connected.
(right panel) An example of a Gabriel graph with $n=100$ nodes.}
\label{Gabriel graph}
\end{figure}

In this section we study the influence of order and randomness on
the melting phase transition in graphs. The influence of order vs.
randomness is on the basis of many physical problems. In particular,
here we are interested in using graphs as a model of solids and granular
materials with ordered structures vs. those having a random one. The
classical example of an ordered system is a crystal where atoms or
molecules are arranged in a repeating pattern \citep{vippagunta2001crystalline}.
On the other hand, amorphous solids, which are characterized by the
lack of regular pattern or repetition, are good examples of random-like
materials \citep{alexander1998amorphous}. In order to model a random-like
material we consider here \textcolor{black}{a type of random graph
known as the Gabriel graph \citep{gabriel1969new}. The Gabriel graphs
$\Gamma=(V,E)$ are constructed by placing randomly and independently
$n$ points in a unit square, then for each pair of points $i\in V$,
$j\in V$, $i\neq j,$ constructs a disk in which the line segment
$\overline{ij}$ is a diameter (see Fig. \ref{Gabriel graph} (left
panel)). The two points will be connected if the corresponding disk
does not contain any other element of $V$. An example of Gabriel
graph is given in Fig. \ref{Gabriel graph} (Right panel). }

%

The reason why we consider random neighborhood (Gabriel) graphs here
instead of other types of random graphs is the following. To keep
the analogy with solid materials we should maintain certain geometric
disposition of the nodes. This geometric arrangements of nodes are
possible in the so-called random geometric graphs (RGGs) as well as
in the random neighborhood graphs (RNGs). The RGGs are nonplanar graphs,
which implies that node $A$ can interact with another $B$ even in
the case that a third node $C$ is exactly in the middle between $A$
and $B$. This, of course, is not a realistic scenario for the interaction
between atoms or molecules and not appropriate for representing a
solid. However, Gabriel graphs are planar graphs and avoid the interaction
between mutually occluded nodes. Consequently, they are appropriate
to model amorphous solids. 

The differences in the ordered vs. random arrangement of atoms/molecules
in crystalline and amorphous solids make that they differ significantly
in the way they change their phase from solid to liquid. That is,
a fundamental difference between crystalline and amorphous solids
resides in the way they melt. While a crystalline solid has a sharp
transition from solid to liquid, the amorphous solid does not. Instead,
it displays a very smooth transition for a long range of temperatures.
The second characteristic feature is that for the same material in
amorphous and crystalline forms, the amorphous one melts at higher
temperature than the crystalline one. For instance, crystalline quartz
melts at $1,550\textdegree C$ and amorphous quartz melts in the range
$1,500-2000\textdegree C$. We are interested in investigating here
this physical reality as an analogy for our crystalline and amorphous
graphs. 

\begin{figure}
\begin{centering}
\includegraphics[width=0.65\textwidth]{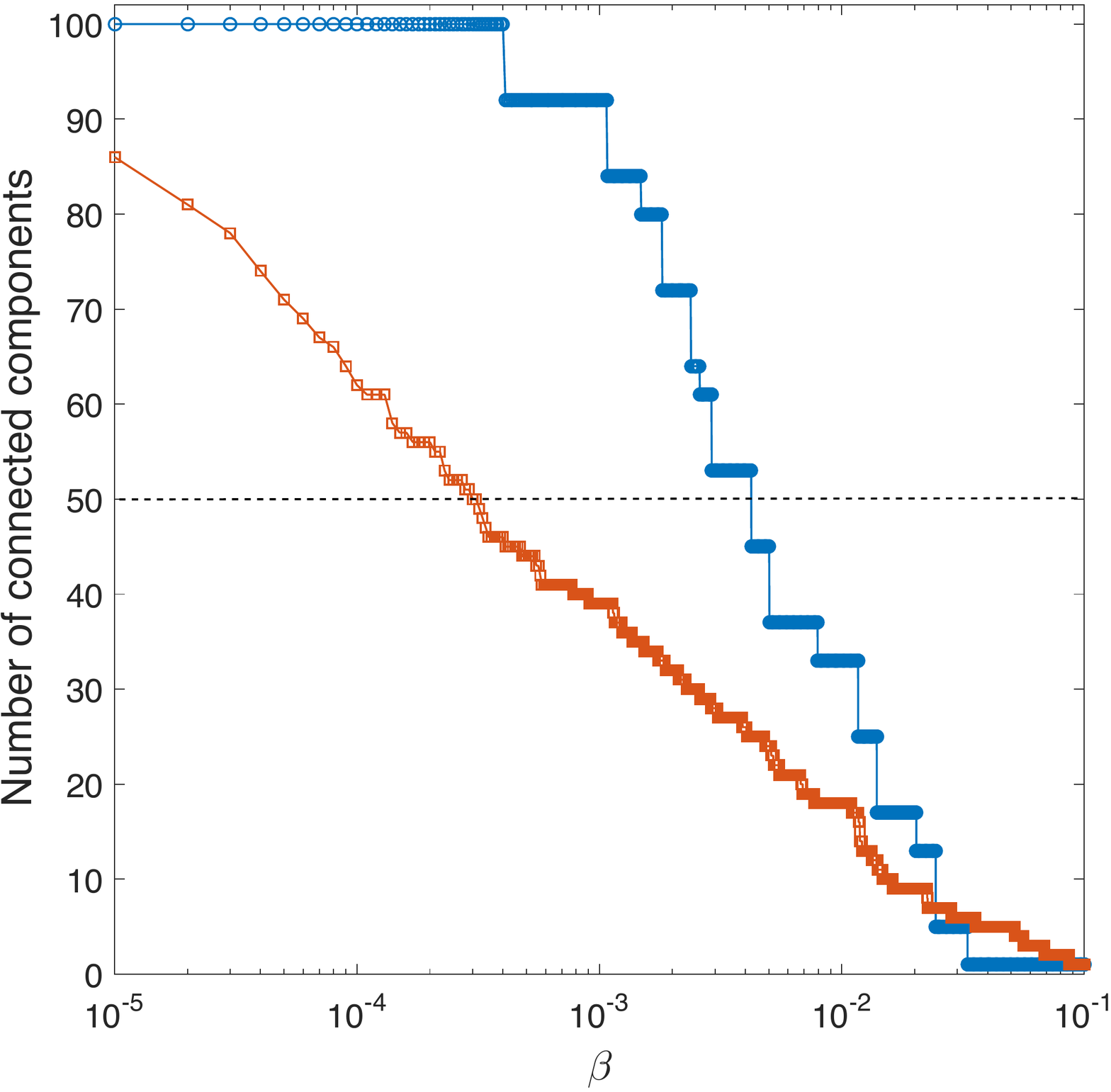}\quad{}
\par\end{centering}
\caption{Change of the number of connected components in the Lindemann graph
for the $10\times10$ square grid (circles) and Gabriel graph with
$n=100$ nodes and $m=180$ edges (squares). The results for the Gabriel
graphs are the average of 100 random realizations. }

\label{Evolution-1}
\end{figure}

In Fig. \ref{Evolution-1} we illustrate the plot of the change in
the number of connected components in the communicability graph with
the change of $\beta$ for a $10\times10$ square grid and a Gabriel
graph with $n=100$ nodes and $m=180$ edges. Similar results to the
ones presented here were obtained for triangular and hexagonal lattices
(results not shown). That is, the main difference between these two
kinds of graphs resides only in the order/randomness of the nodes
in a unit square. The lattice representing a crystalline graph and
the Gabriel graph representing an amorphous one. The main observation
is that while the crystalline graph displays a sharp increase in the
number of connected components with the decrease of $\beta$, the
amorphous one displays a rather slow change (notice that the $x$-axis
is in logarithmic scale). The second important observation is that
the structure of the crystalline graph is destroyed more quickly than
that of the amorphous one. For instance, if we consider the value
of $\beta$ at which the number of connected components is exactly
half the number of nodes, we can see that the crystalline graph reaches
that point an order of magnitude before than the amorphous one. If
we identify this value of $\beta$ as the melting temperature of a
graph we can say that the crystalline graph has a melting temperature
one order of magnitude smaller than that of an amorphous one. We repeated
the experiment with a $25\times25$ square lattice and the corresponding
Gabriel graph to see whether there are some small size effects and
observed that the results are very much the same, although we have
increased the size of the graphs by a factor of 6 (results not shown).

Another possibility of the current approach is that it allows us to
visualize the evolution of the ``melting'' process in graphs in
order to gain insights about its mechanism. In Fig. \ref{Evolution-1}
we illustrate some snapshots of the change in the communicability
structure with the change of $\beta$ for the square lattice. We represent
in red the nodes that have removed all of their edges and are now
disconnected from the giant connected component. In blue we represent
those nodes which form the giant connected component of the graph.

\begin{figure}
\subfloat[]{\begin{centering}
\includegraphics[width=0.3\textwidth]{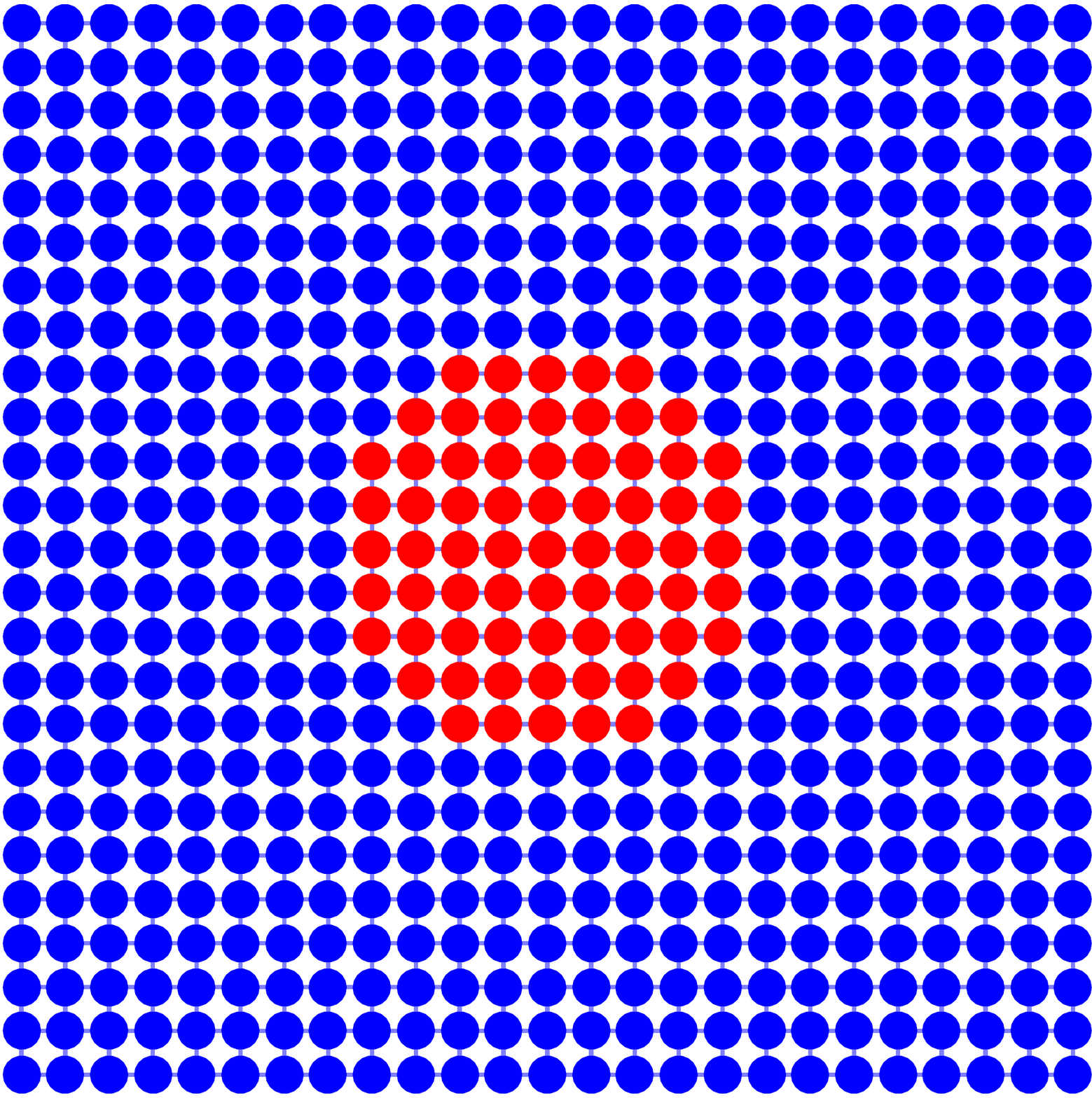}
\par\end{centering}

}\subfloat[]{\begin{centering}
\includegraphics[width=0.3\textwidth]{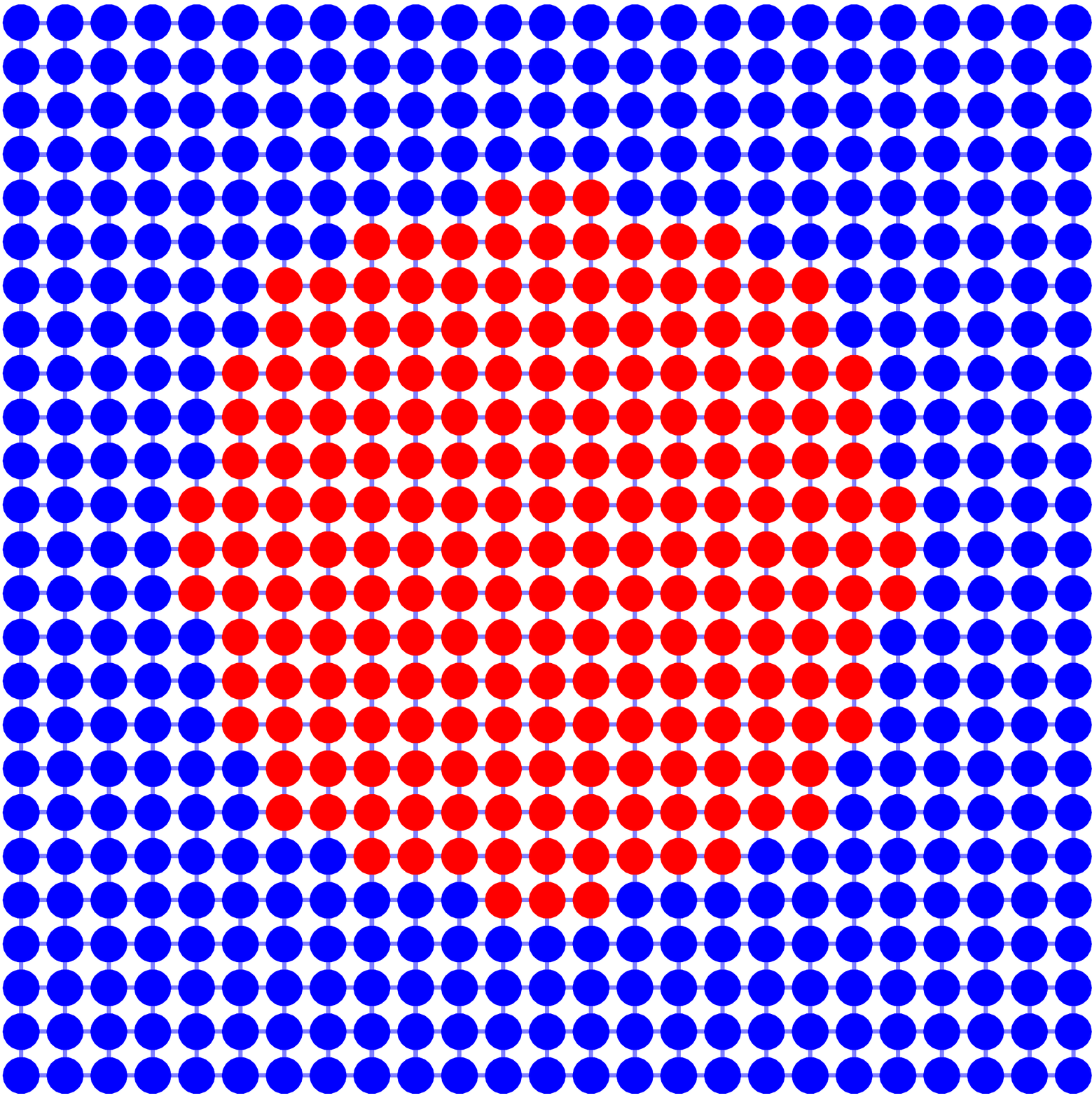}
\par\end{centering}
}\subfloat[]{\begin{centering}
\includegraphics[width=0.3\textwidth]{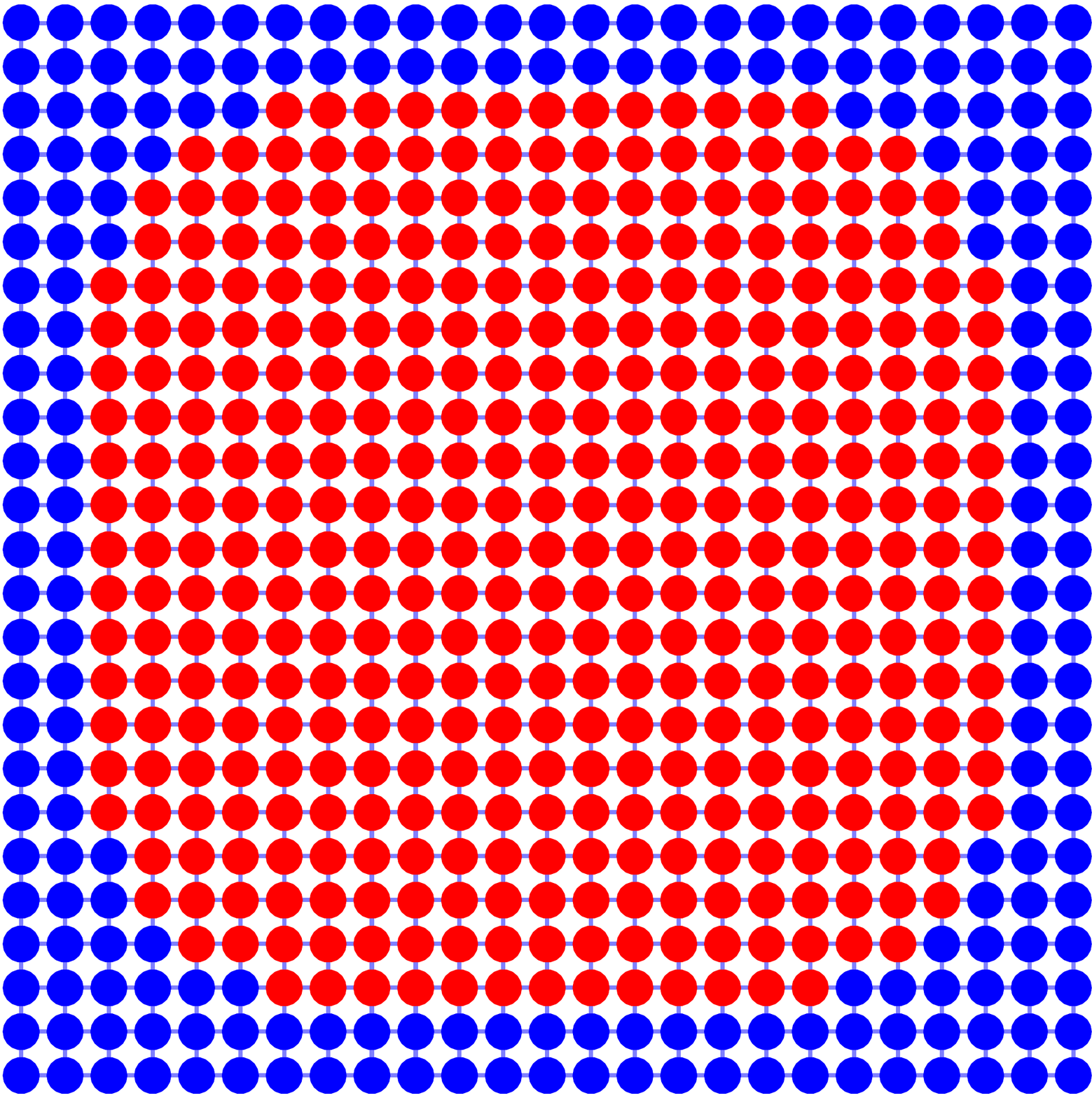}
\par\end{centering}
}
\begin{centering}
\subfloat[]{\begin{centering}
\includegraphics[width=0.7\textwidth,angle=90]{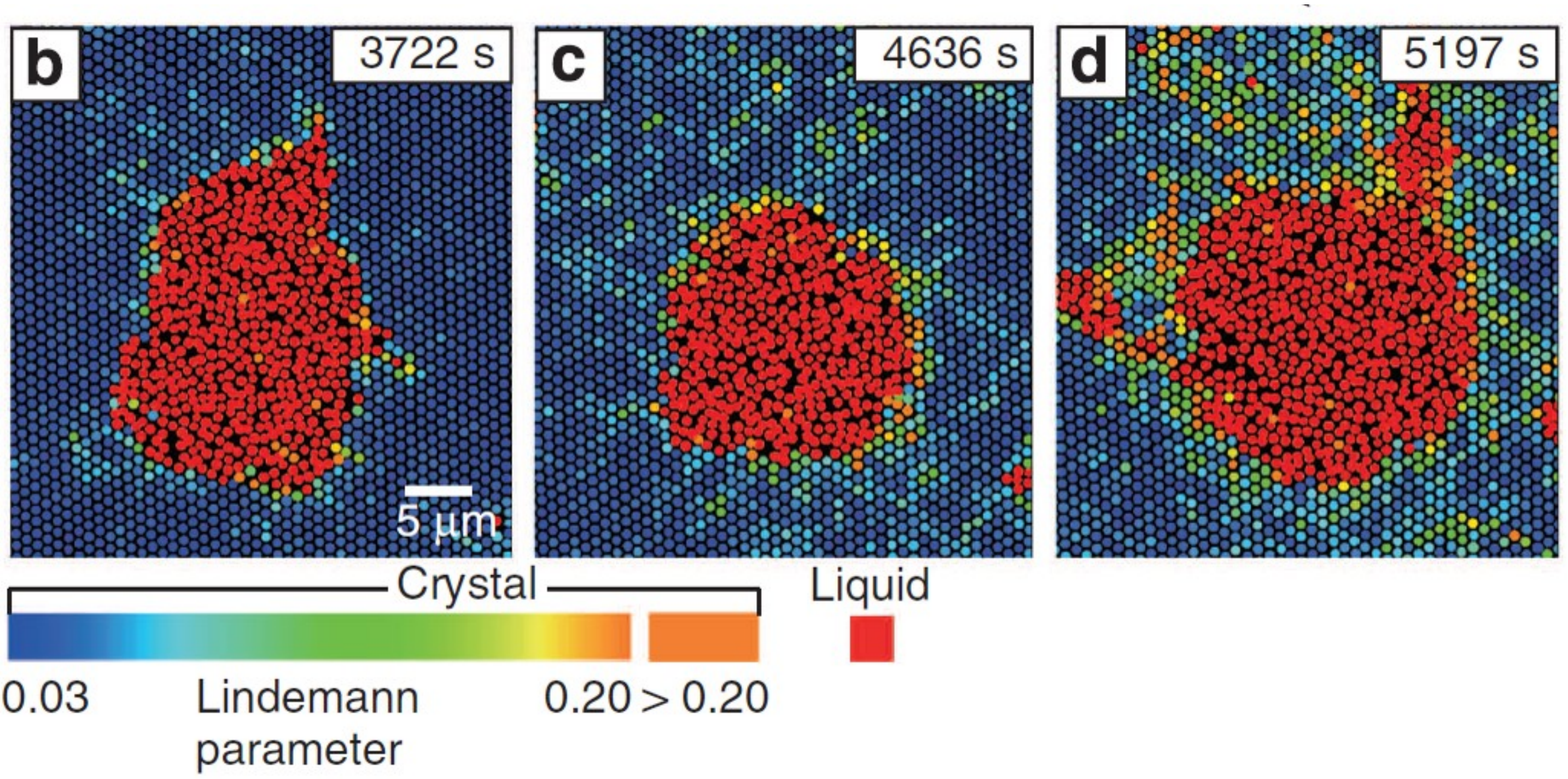}
\par\end{centering}
}
\par\end{centering}
\caption{Illustration of the melting of a $25\times25$ square lattice at $\beta=0.000075$
(a), $\beta=0.00005$ (b) and $\beta=0.000025$ (c). Results for the
melting of colloidal crystals obtained by Wang et al. \citep{wang2015direct}.
In plots (a)-(c) the nodes not in the giant connected component are
colored in red. ~}

\label{eigenvector}
\end{figure}

At very low values of $\beta$, e.g. $\beta=0.000025$ (Fig. \ref{eigenvector}
(c)) the communicability structure of the lattice resembles a trivial
graph in which almost every node is isolated. As the temperature drops,
$\beta$ increases, certain structures start to emerge. In particular,
for $\beta=0.0005$ (Fig. \ref{eigenvector} (b)) an annulus\textendash external
part of the lattice\textendash is solidified into a single connected
component and only the central part of the graph remains melted. As
the temperature drops below $\beta=0.000075$ (Fig. \ref{eigenvector}
(a)), the melted region\textendash red nodes\textendash shrinks to
the very center of the lattice. The observed pattern of melting of
the square lattice is similar to the one observed experimentally for
crystalline solids. In Fig. (Fig. \ref{eigenvector} (d)) we illustrate
the results of Wang et al. \citep{wang2015direct} for the melting
of colloidal crystals which show such patter of central melting. 

\begin{figure}
\begin{centering}
\includegraphics[width=0.3\textwidth]{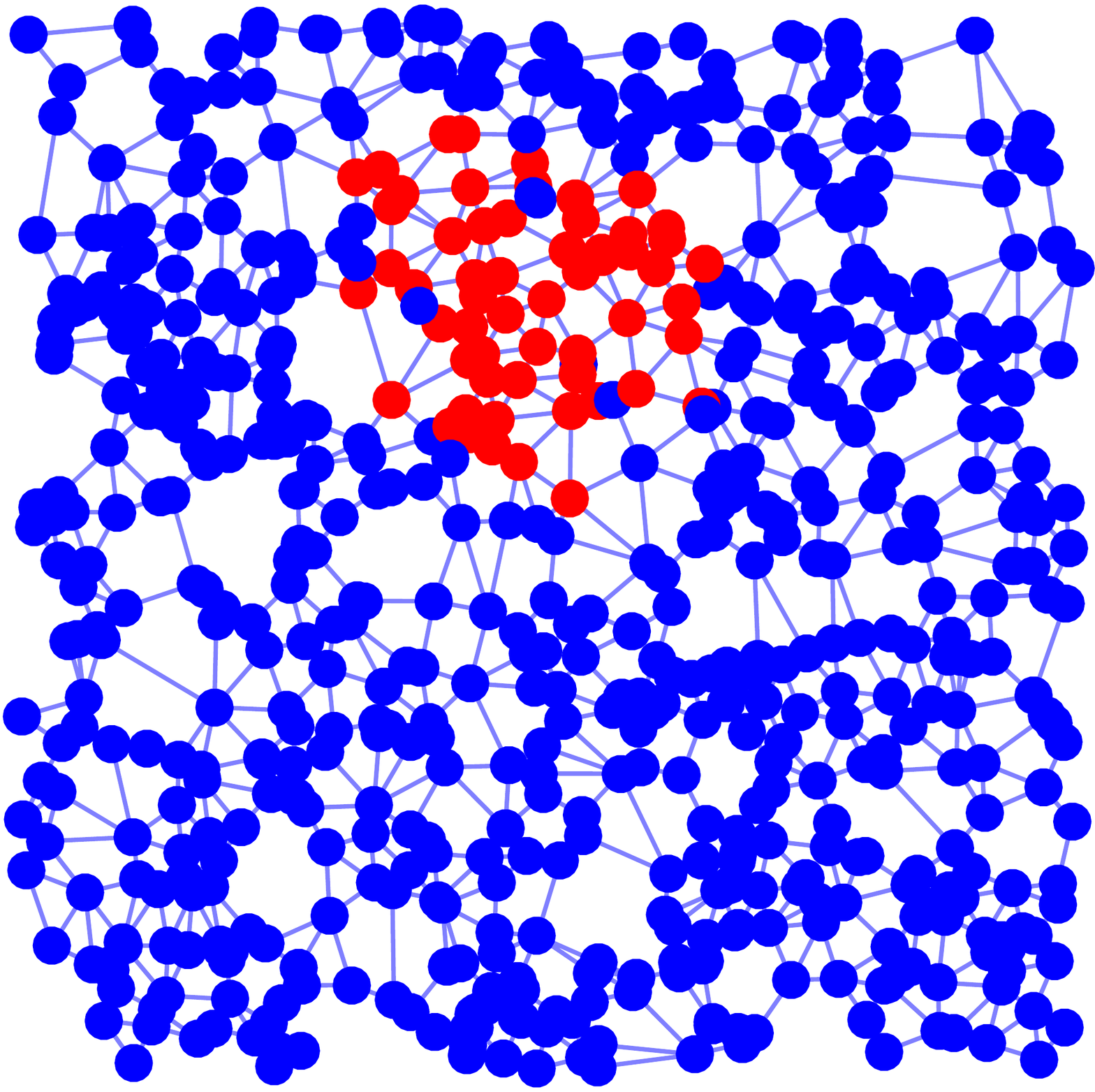}\quad{}\includegraphics[width=0.3\textwidth]{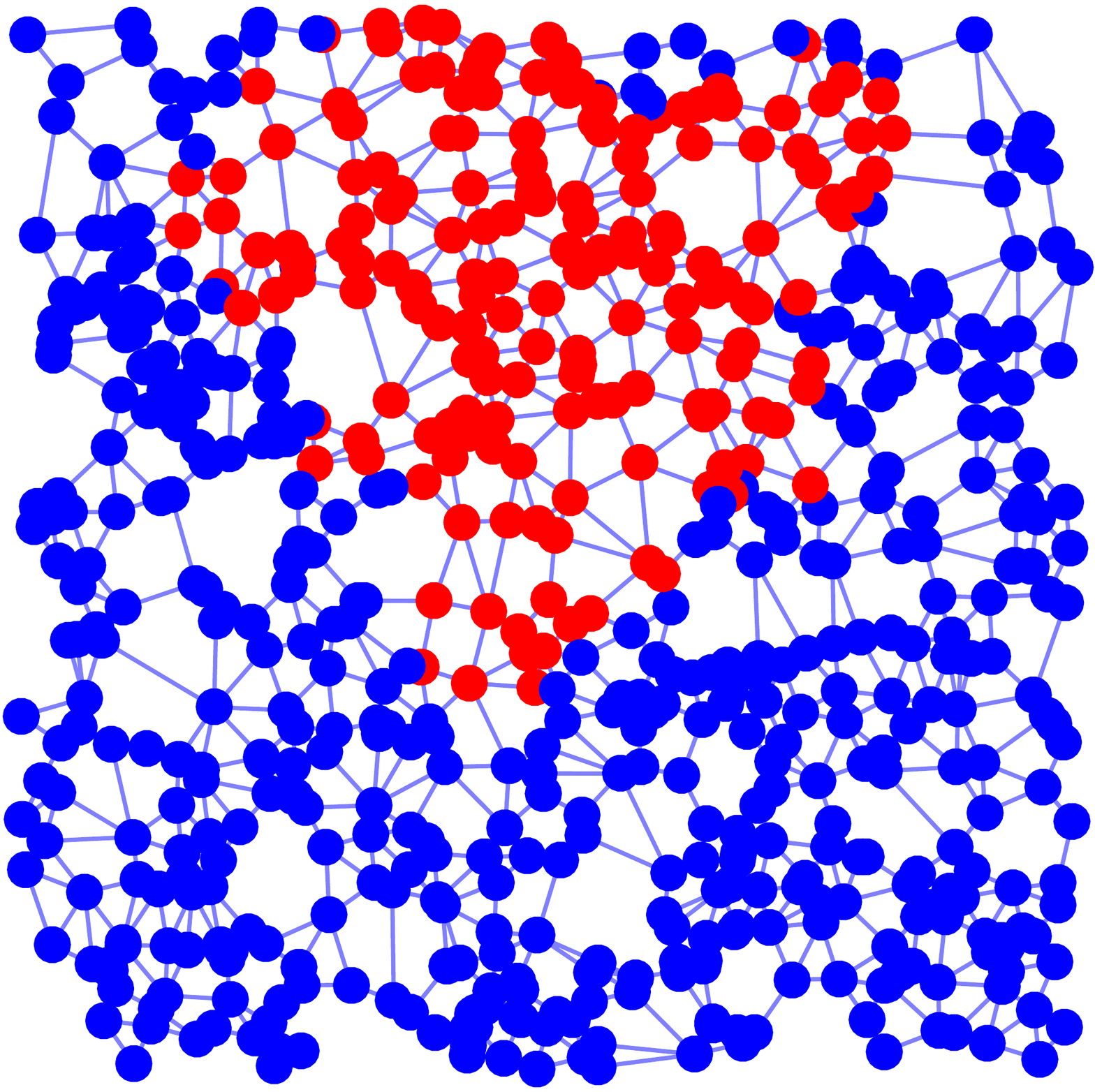}\quad{}\includegraphics[width=0.3\textwidth]{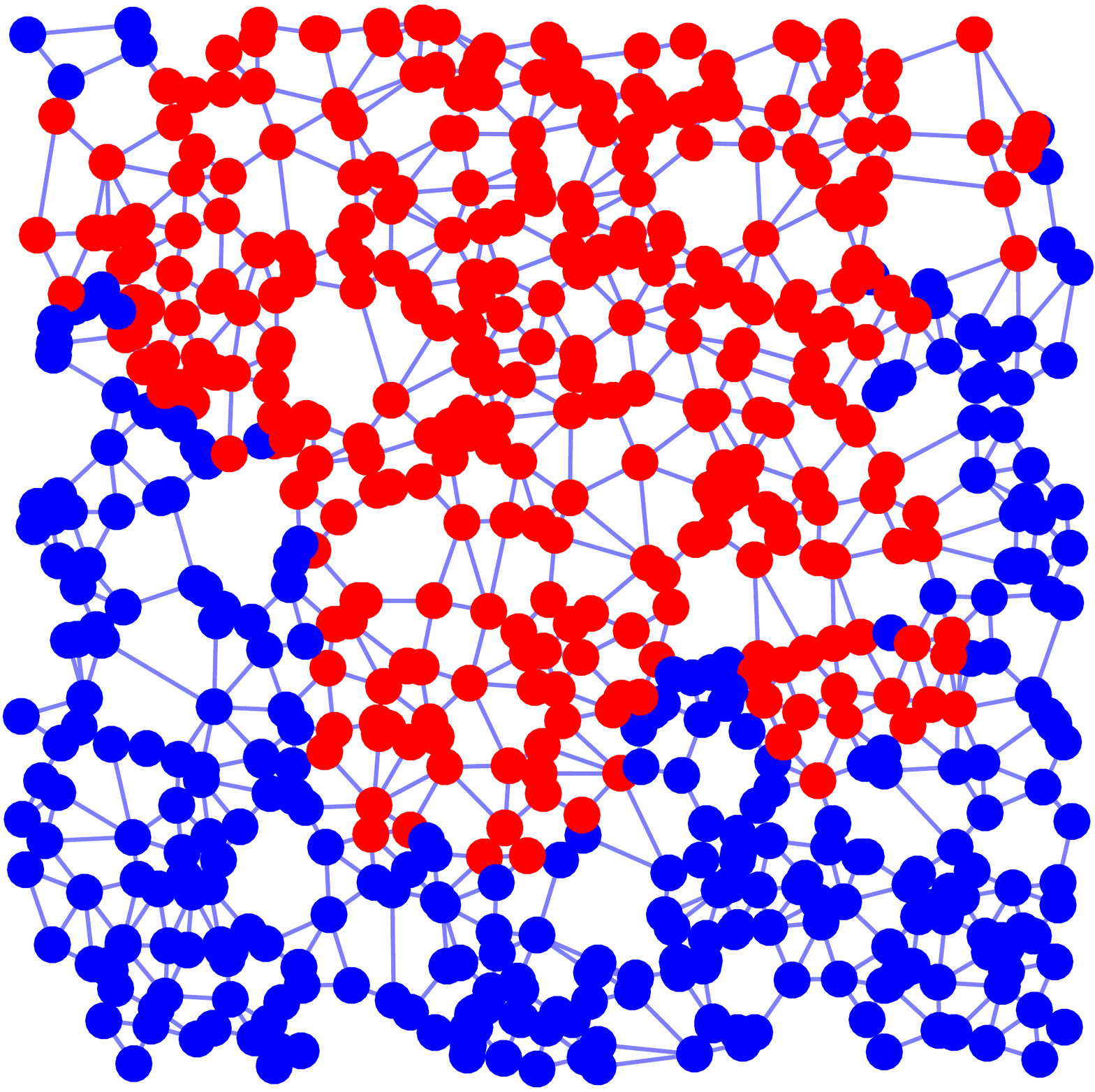}
\par\end{centering}
\caption{Nodes in each of the connected components of the Gabriel graph studied
here for $\beta=10^{-8}$, $10^{-9}$ and $10^{-10}$ from left to
right. The nodes not in the giant connected component are colored
in red. }

\label{Evolution}
\end{figure}

In the case of the amorphous graph there is no repeating pattern in
them, and it is impossible to find a general structural pattern of
the evolution of the melting process. A few snapshots of the process
are given in Fig. \ref{Evolution}. The temperature needed to melt
these graphs is significantly higher\textendash smaller $\beta$\textendash than
the ones needed to melt square lattices of the same size, which coincides
with our previous observations as well as with the experimental results
for crystalline and amorphous solids. The reasons of this significant
difference will become clear in the next section of this work.

\subsection{Complex networks}

The term complex networks is frequently used to refer to graphs representing
the skeleton of complex systems, such as social, ecological, infrastructural,
technological and biomolecular networks \citep{estrada2012structure}.
Here we consider a series of 47 complex networks\textendash see Supplementary
Information for description and references\textendash arising from
these different scenarios. Here we split our analysis into two parts.
First we consider global properties of the networks and then we analyze
the influence of node-level centrality on the melting process of these
networks. 

\subsubsection{Global analysis}

We start here by finding the value of $\beta$ at which the transition
between connected to disconnected Lindemann graph occurs. Our first
task is to relate the values of $\beta_{c}$ to some simple topological
parameters of the networks in order to understand the structural dependence
of this transition. With this goal we study the following structural
representative parameters of networks: edge density $\delta$, average
degree $\bar{k}$, maximum degree $k_{max}$, average Watts-Strogatz
clustering coefficient $\bar{C}$, average path length $\bar{l}$,
shortest path efficiency $E$, spectral radius of the adjacency matrix
$\lambda_{1}$, second largest eigenvalue of the adjacency matrix
$\lambda_{2}$, spectral gap of the adjacency matrix $\Delta$, average
communicability distance $\bar{\xi}$, average resistance distance
$\bar{\Omega}$, and average communicability angles $\bar{\theta}$.
The definitions of these measures are given in the Supplementary Information
accompanying this work. We investigate correlations between these
measures and the values of $\beta_{c}$ for the 47 networks studied
here in linear, semi-log and logarithmic scales. The most significant
correlation was obtained for $\ln\beta_{c}$ and $\ln\delta$ ($r=0.79),$
where $r$ is the Pearson correlation coefficient. Also significant
are the correlations between $\ln\beta_{c}$ and $\bar{l}$ ($r=-0.72),$
and with $\ln E$ ($r=0.72$). 

The correlations found for $\ln\beta_{c}$ with some of the previous
structural parameters may be hiding something about the real structural
characteristic of networks that influence their ``melting''. For
instance, the negative correlation between edge density and $\beta_{c}$
seems suspicious. Our intuition tells us that, under all other structural
conditions the same, high density networks should melt at higher temperatures,
i.e., lower $\beta_{c}$, than lower density ones. This is exactly
what it is observed in molecular crystals of nonpolar molecules, such
as linear alkanes \citep{boese1999melting}. In addition, in the previous
section we have studied two different kinds of networks which differ
very significantly in theirs values of $\beta_{c}$ in spite of the
fact that they have exactly the same number of nodes and edge densities.
While the square lattice is an almost regular graph, in a Gabriel
graph there are small degree heterogeneities that emerge from the
clustering of groups of nodes in a relatively close space. Then, the
fact that smaller and denser (real-world) networks are the ones having
the largest $\beta_{c}$, i.e., they have Lindemann graphs easier
to disconnect, may indicate that the ``degree homogeneity'' of these
networks more than their sizes or densities is the real driver of
their melting. In order to capture these degree irregularities we
recall the definition of the average degree of a network

\begin{equation}
\bar{k}=\dfrac{2m}{n}=\dfrac{\vec{1}^{T}A\vec{1}}{\vec{1}^{T}\vec{1}}.
\end{equation}
The right-hand side of the previous equation is useful to think that
the spectral radius of the adjacency matrix is a sort of average degree,
which instead of counting only the number of nearest neighbors of
a node consider also a more global picture around it

\begin{equation}
\lambda_{1}=\dfrac{\vec{\psi}_{1}^{T}A\vec{\psi}_{1}}{\vec{\psi}_{1}^{T}\vec{\psi}_{1}}.
\end{equation}

Notice that $\bar{k}\leq\lambda_{1}$ with equality if and only if
the graph is regular. Thus, the term $\left(\lambda_{1}/\bar{k}\right)$
represents the ratio of a more global environment of a node to its
more local one. That is, the ratio $\left(\lambda_{1}/\bar{k}\right)$
indicates how a node ``sees'' as average its global environment
in relation to its nearest neighbors. In a regular graph its local
environment, i.e., its degree, is identical to the degree of its neighbors,
second neighbors, and so on and we get that $\left(\lambda_{1}/\bar{k}\right)$=1.
Then, we can define the following index of global to local degree
heterogeneity 

\begin{equation}
\varrho\left(G\right)=n\left(\dfrac{\lambda_{1}}{\bar{k}}\right).
\end{equation}

\begin{figure}
\begin{centering}
\includegraphics[width=0.8\textwidth]{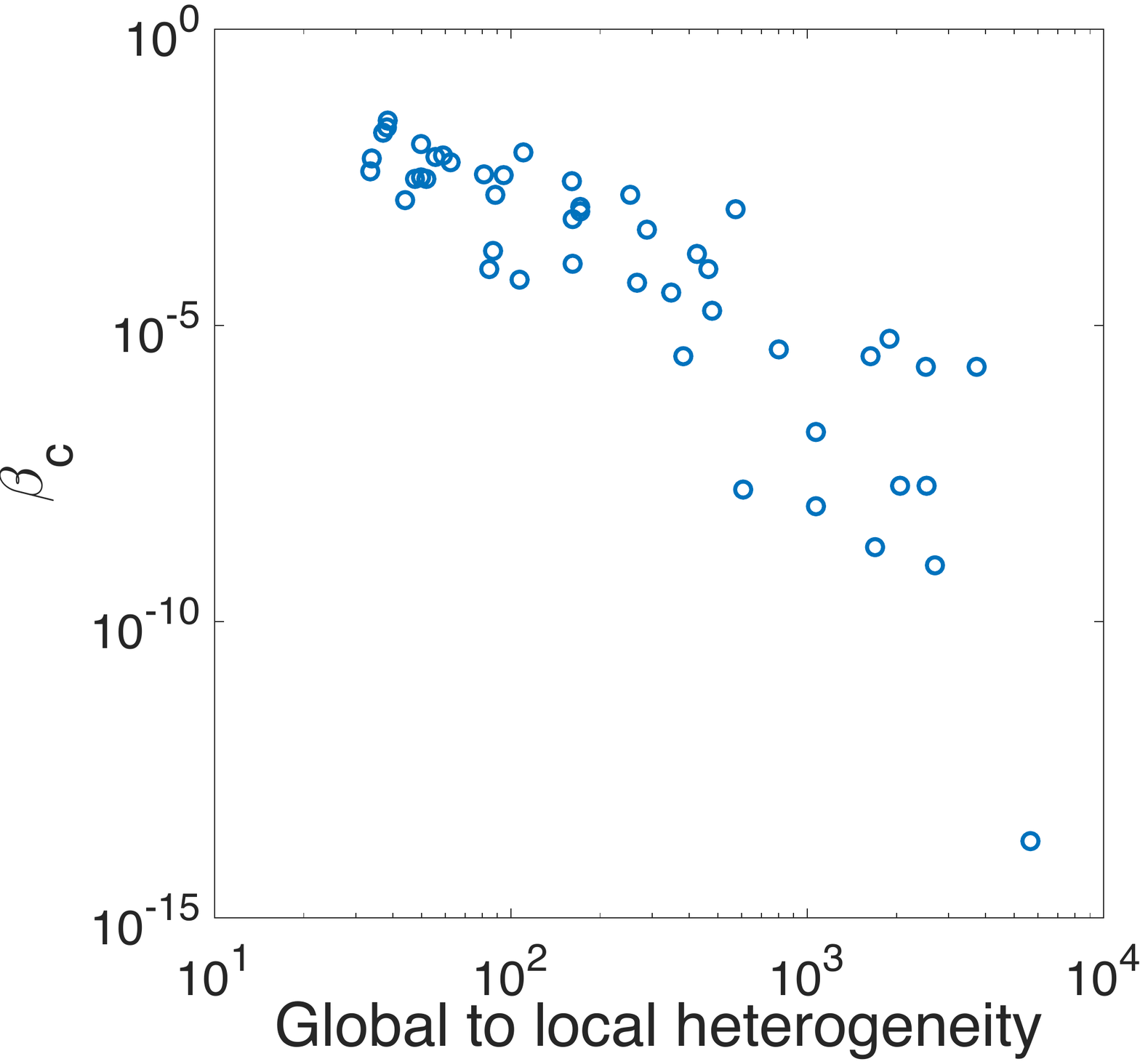}
\par\end{centering}
\caption{Changes of $\beta_{c}$ for 47 real-world networks as a function of
their global/local degree heterogeneity as described in this work. }

\label{heterogeneity}
\end{figure}

Notice that $\varrho\left(G\right)=\left(\dfrac{\lambda_{1}}{\delta}\right)$,
which may explain the previously observed correlation between $\ln\beta_{c}$
and $\ln\delta$. We have then used $\varrho\left(G\right)$ as an
indicator of the global to local heterogeneity of the 47 real-world
networks analyzed here. In Fig. \ref{heterogeneity} we illustrate
the log-log plot of $\varrho\left(G\right)$ versus $\beta_{c}$,
which has correlation coefficient $r=-0.85$. The values of $\varrho\left(G\right)$
also explain the differences in the melting of the square lattice
($\varrho\left(G\right)\approx641.89$) and the Gabriel graphs ($\varrho\left(G\right)\approx780.62$)
studied in the previous section. 

The most important message of this section is the following. The disconnection
of the Lindemann graph of a given graph, i.e., its melting, depends
very much on the differences between global and local degree heterogeneities.
Regular graphs are easier to melt than nonregular ones, and the more
irregular\textendash in terms of global to local degree heterogeneity\textendash the
graph is the smallest the value of $\beta_{c}$, i.e., more difficult
to melt. 

\subsubsection{Local analysis }

\begin{figure}
\includegraphics[width=0.3\textwidth]{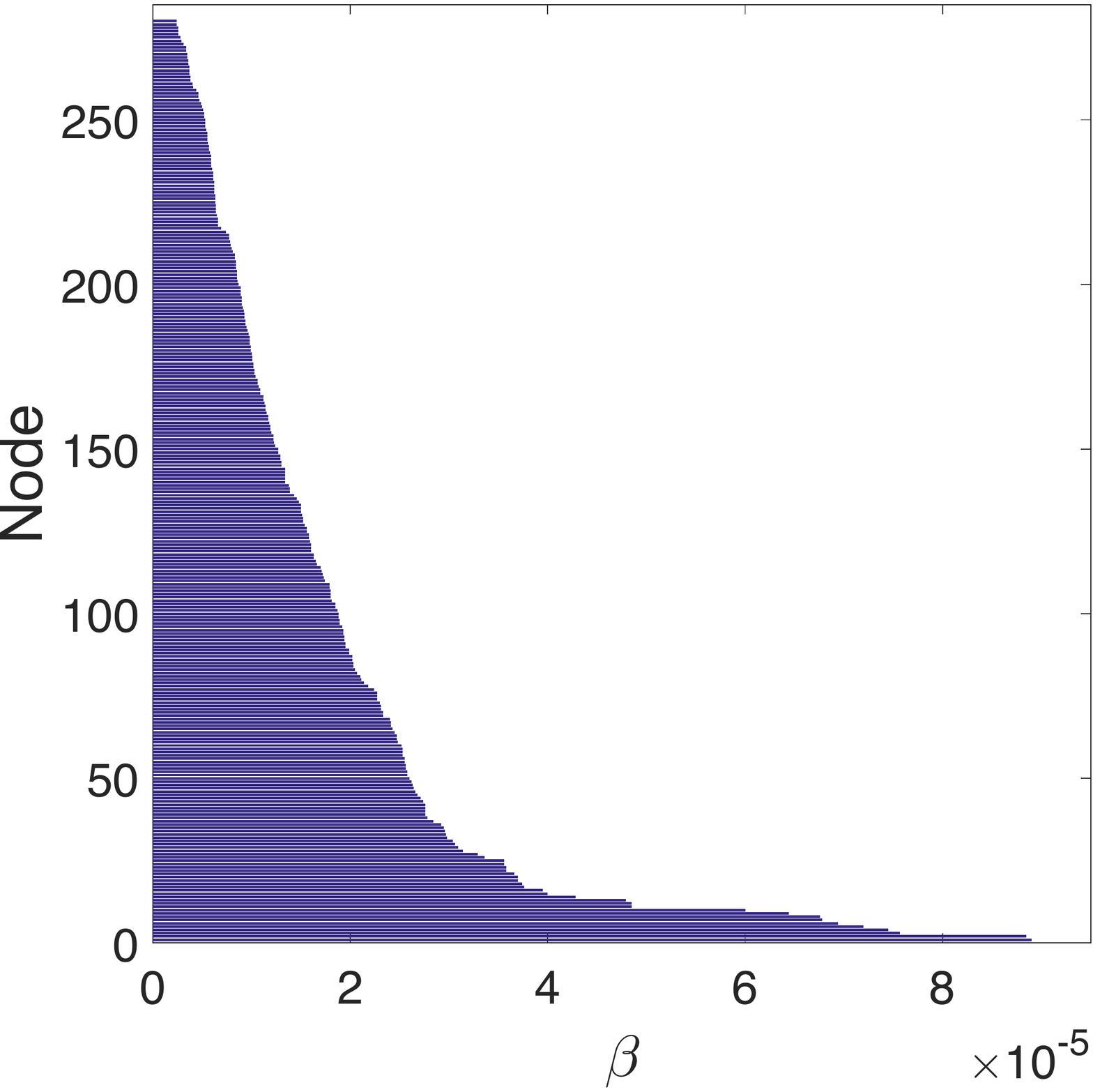}
\includegraphics[width=0.3\textwidth]{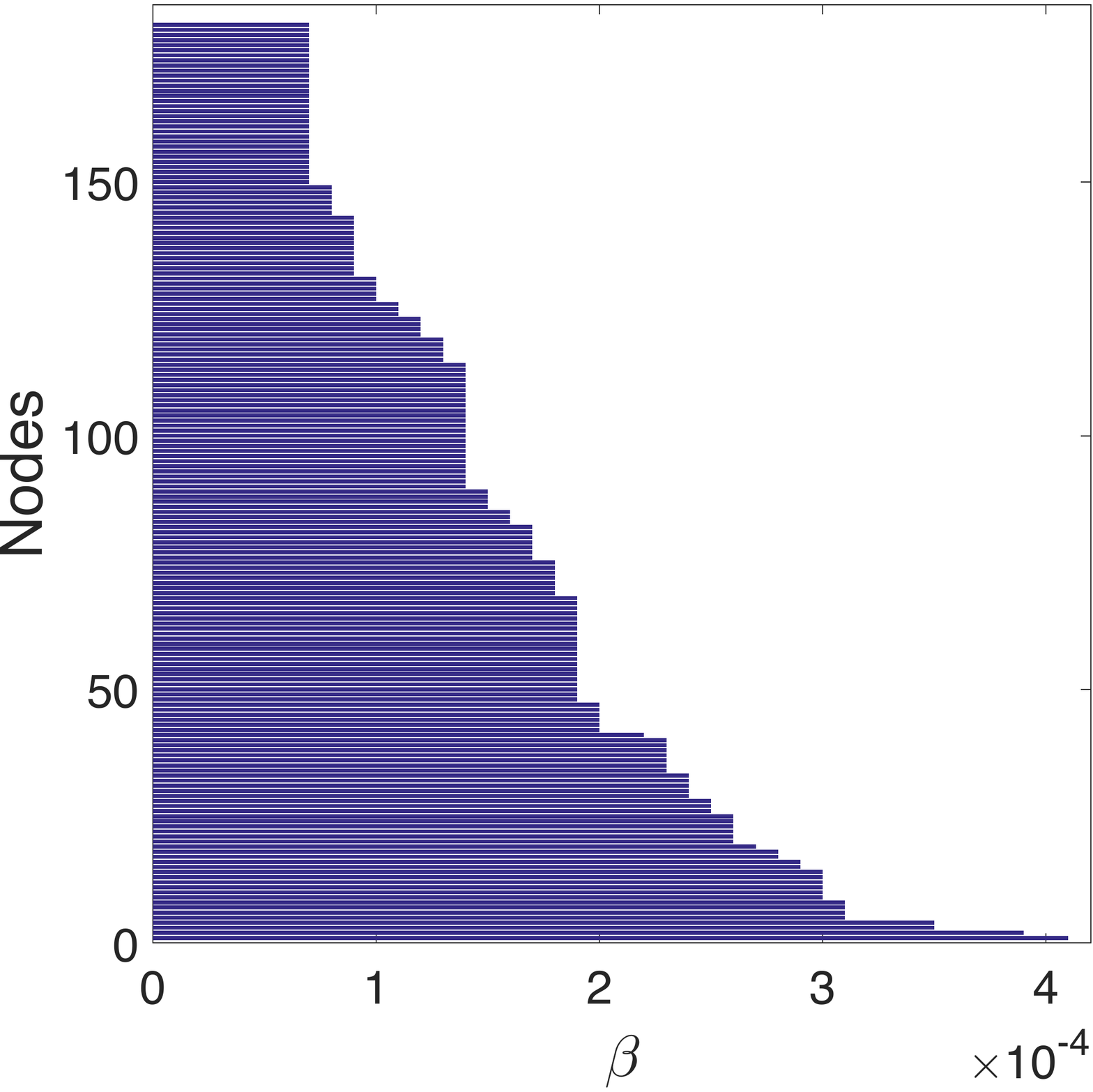}
\includegraphics[width=0.3\textwidth]{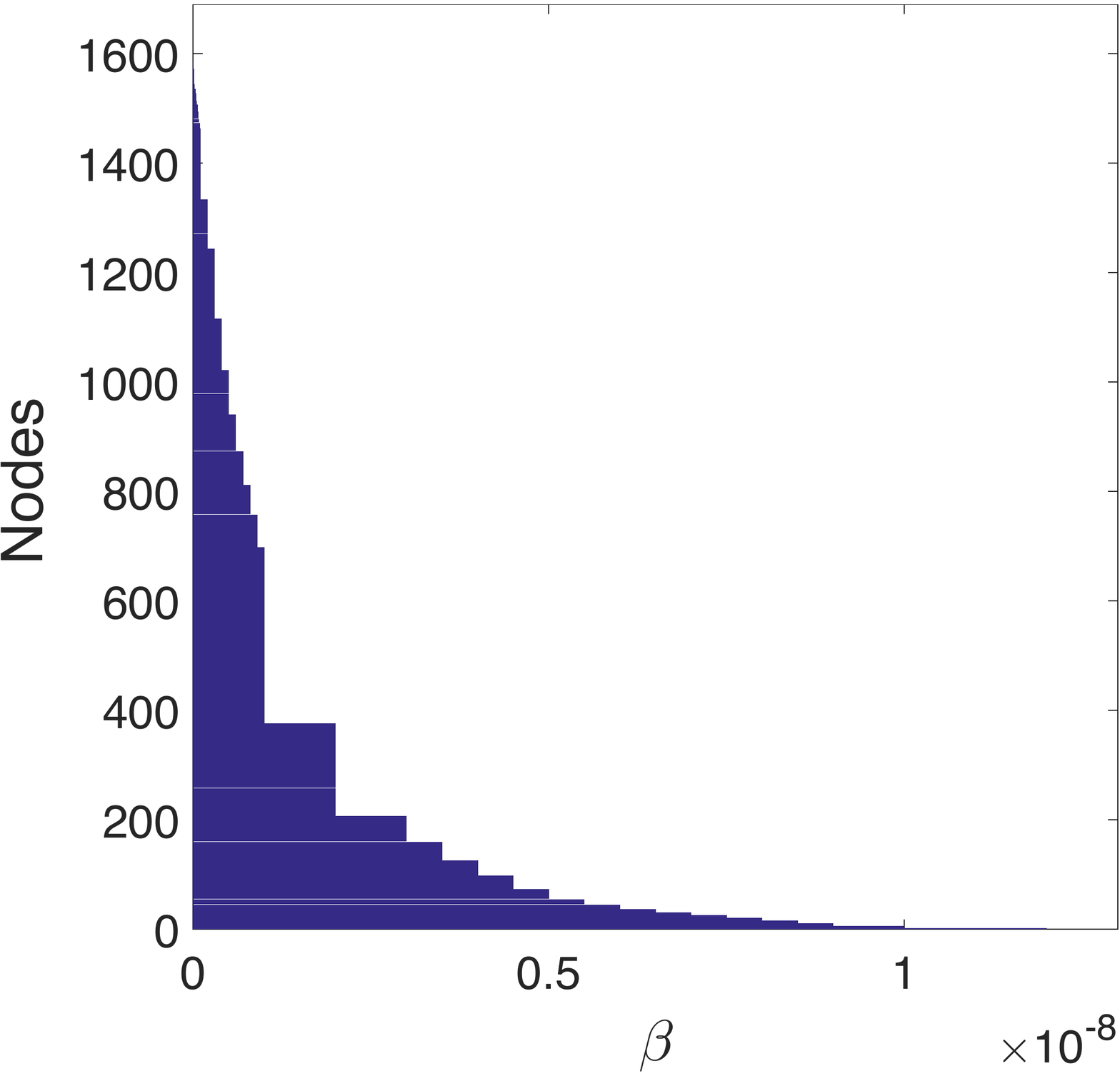}
\caption{Illustration of the melting barcodes of the networks of neurons (a),
Little Rock (b) and corporate elite (c).}

\label{barcodes}
\end{figure}

In this subsection we are interested in the local analysis of the
effects of decreasing the value of $\beta$ on the topological structure
of a network. In particular we investigate computationally two important
aspects of the graph melting process: (i) How the nodes of a network
melt? and (ii) Which structural parameter drives the melting of the
nodes? For investigating these questions we consider a subset of the
real-world networks studied in this work. We create a melting barcode
plot in which we plot every node in the $y$-axis and in the $x$-axis
we provide the value of $\beta$ at which the corresponding node disconnect
from the giant connected component of the graph. In Fig. \ref{barcodes}
we illustrate the melting barcodes of three networks: neurons (a),
Little Rock (b) and corporate elite (c). We need to read these melting
barcodes from right to left as the melting process starts at higher
values of $\beta$ and proceed by decreasing it. There are significant
differences in the three barcodes presented which point out to the
differences existing in the melting processes of the different graphs
analyzed. First, we can observe that the shape of the melting barcodes
are different. While in ``neurons'' the decay resembles an exponential
curve, in ``Little Rock'' it is almost linear and in the ``corporate
elite'' it displays a more skewed shape (see further for quantitative
analysis). In the second place, the barcodes of ``Little Rock''
and of corporate elite display regions in which large groups of nodes
are disconnected at the same temperature, while in neurons the change
is smoother.

We then investigate the rate of change of the melting process in the
networks analyzed by considering the shape of the histogram of the
number of nodes ``melted'' at a given temperature. That is, we construct
the histograms of the number of nodes melted in a temperature range
versus the range of temperatures. In general we observe two kinds
of decay of the number of nodes melted at a given temperature in relation
to the inverse temperature. They are:

\begin{equation}
\eta=a\exp\left(\zeta\cdot\beta\right),\label{eq:exponential}
\end{equation}

\begin{equation}
\eta=b\cdot\beta^{\gamma},\label{eq:power_law}
\end{equation}
where $\eta$ is the number of nodes melted at a given value of $\beta$.
For some of the smallest networks it was not possible to find any
particular law of the decay of $\eta$ as a function of $\beta$.
These were the cases of the networks of Benguela ($n=29$), Coachella
($n=30$), Social3 ($n=32$), St. Marks ($n=48$), as well as for
the network of Little Rock, which is not so small ($n=181$) but it
also has a very disperse histogram. For the rest of the networks analyzed
we display the parameters of the fitting to Eqs. \ref{eq:exponential}
and \ref{eq:power_law} in Table \ref{Table_fitting}.

\begin{table}
\begin{centering}
\begin{tabular}{|c|c|c|c||c|c|c|c|}
\hline 
\multirow{2}{*}{network} & \multicolumn{3}{c||}{Eq. \ref{eq:exponential}} & \multirow{2}{*}{network} & \multicolumn{3}{c|}{Eq. \ref{eq:power_law}}\tabularnewline
\cline{2-4} \cline{6-8} 
 & $a$ & $\zeta$ & $r^{2}$ &  & $b$ & $\gamma$ & $r^{2}$\tabularnewline
\hline 
\hline 
Prison & $24.58$ & $-1.858\cdot10^{-3}$ & $0.790$ & Macaque & $5.59\cdot10^{-15}$ & $-6.953$ & $0.756$\tabularnewline
\hline 
Neurons & $178$ & $-6.032\cdot10^{-4}$ & $0.975$ & Stony & $4.84\cdot10^{-12}$ & $-3.515$ & $0.887$\tabularnewline
\hline 
Small World & $334.9$ & $-1.06\cdot10^{-4}$ & $0.995$ & PIN \textit{B. subtilis} & $2.34\cdot10^{-3}$ & $-1.016$ & $0.976$\tabularnewline
\hline 
Ythan & $91.82$ & $-3.61\cdot10^{-3}$ & $0.965$ & Roget & $5.40\cdot10^{-8}$ & $-1.441$ & $0.992$\tabularnewline
\hline 
Electronic 1 & $75.87$ & $-5.99\cdot10^{-4}$ & $0.934$ & Software\_Abi & $7.05\cdot10^{-17}$ & $-2.621$ & $0.999$\tabularnewline
\hline 
PIN \textit{H. pylori} & $1233$ & $-2.89\cdot10^{-6}$ & $0.999$ & Corporate elite & $9.96\cdot10^{-14}$ & $-1.744$ & $0.999$\tabularnewline
\hline 
\end{tabular}
\par\end{centering}
\caption{Values of the fitting parameters for the Eqs. \ref{eq:exponential}
and \ref{eq:power_law} displaying the relation between the number
of nodes melted at a given value of $\beta$ as a function of $\beta$
for several real-world networks.}

\label{Table_fitting}
\end{table}

The fitting parameters given in Table \ref{Table_fitting} indicate
the differences in the rates of melting of the networks analyzed.
These rates of melting represent a new measure of the robustness of
networks to the effects of external stresses to which the networks
are submitted to, as accounted for by the inverse temperature. For
instance, those networks melting according to Eq. \ref{eq:exponential}
are more robust to external stresses than the ones melting according
to Eq. \ref{eq:power_law}. In comparing those networks that melt
exponentially with $\beta$ it is clear that the social network of
inmates in prison (Prison) and the food web of Ythan are significantly
less robust to such external stresses than the protein interaction
network of \textit{H. pylori. }The network representing the visual
cortex of macaque melts very quickly in relation to the rest of the
networks analyzed indicating that once the external stress has trigger
the melting process the nodes of this network disconnect very fast
from the giant connected component. 

Finally, we investigate which structural parameters determine the
melting process of the nodes of a network. In particular we consider
here the role of node centrality on the melting of the corresponding
node. We then analyze the relation between the value of $\beta$ at
which a node melts and its degree (DC), closeness (CC), betweenness
(BC), eigenvector (EC) and subgraph centrality (SC). All these measures
are defined in the Supplementary Information. In general, we observe
that the values of $\beta$ at which the nodes melt correlate very
well with EC. All networks studied displayed Pearson correlation coefficients
between these two parameters higher than $0.90$, with the exceptions
of the networks of Benguela and Macaque visual cortex. In addition
we investigate the coefficient of variation (CV) of the values of
$\beta$ at which a node melts estimated from a linear regression
with EC. This coefficient is given by the standard deviation of the
estimate divided by the mean of the values of $\beta$ at which the
nodes melt in a given network. Here we provide the values of both
Pearson correlation coefficient and CV in percentage for the networks
investigated: Benguela ($r=0.68$, 34.3\%), Coachela ($r=0.93$, 15.1\%),
Social3 ($r=0.95$, 16.2\%), Macaque ($r=0.82$, 19.8\%), St. Marks
($r=0.97$, 11.7\%), Prison ($r=0.998$, 3.9\%), PIN \textit{B. subtilis}
($r=0.999$, 3.4\%), Stony ($r=0.94$, 13.8\%), Electronic1 ($r=0.99999$,
0.4\%), Ythan1 ($r=0.99$, 8.6\%), Small World ($r=0.998$, 5.0\%),
Little Rock ($r=0.989$, 6.9\%), Neurons ($r=0.999$, 1.3\%), Roget
($r=0.997$, 7.6\%), PIN \textit{H. pylori} ($r=0.997$, 7.6\%), Software
Abi ($r=0.996$, 19.7\%), Corporate elite ($r=0.99$, 18.1\%). An
important point to have into account here is that although the correlation
coefficients are in general very high, the values of CV indicate that
the correlations are characterized by certain levels of dispersion.
For instance, the networks of Software Abi and the Corporate elite
have CV close to 20\% although they have correlation coefficients
larger than 0.99. 

In general, we observe that when $\beta_{c}$ is arbitrarily small,
the correlation between the value of $\beta$ at which the node melts
and EC is better than when $\beta_{c}$ is relatively large, e.g.
Benguela, Coachela, Social3, Macaque, St. Marks, etc. The reason for
that difference is the following. Let us recall that at $\beta_{c}$
the value of the Lindemann criterion is negative, that is 

\begin{equation}
\Delta\tilde{G}_{pq}\left(\beta_{c}\right)=\sum_{j=2}^{n}\psi_{j}\left(p\right)\psi_{j}\left(q\right)e^{\beta\lambda_{j}}<0.
\end{equation}

Let $\beta_{c}$ be arbitrarily small such that we have $e^{\beta_{c}\lambda_{j}}\approx1$
for all $j$ and

\begin{equation}
\begin{split}-\left|\Delta\tilde{G}_{pq}\left(\beta_{c}\right)\right| & =M\left(\Gamma,\beta_{c}\right)+\sum_{j=2}^{n}\psi_{j}\left(p\right)\psi_{j}\left(q\right)\\
 & =M\left(\Gamma,\beta_{c}\right)-\psi_{1}\left(p\right)\psi_{1}\left(q\right),
\end{split}
\end{equation}
where $M\left(\Gamma,\beta_{c}\right)$ is obviously a constant. Then,
if we take the sum of all the values of $\Delta\tilde{G}_{pq}\left(\beta_{c}\right)$
for the node $p$ we have

\begin{equation}
\begin{split}-\sum_{q\neq p}\left|\Delta\tilde{G}_{pq}\left(\beta_{c}\right)\right| & =M\left(\Gamma,\beta_{c}\right)-\psi_{1}\left(p\right)\sum_{q\neq p}\psi_{1}\left(q\right)\end{split}
\end{equation}
which clearly explains the observed high positive correlation between
the values of $\beta$ at which a node melts and $\psi_{1}\left(p\right)$
for networks having $\beta_{c}$ very close to zero. Also, it explains
why those networks for which $\beta_{c}$ is not sufficiently small
display bad correlations between the values of $\beta$ at which a
node melts and $\psi_{1}\left(p\right)$. 

\begin{figure}
a)\includegraphics[width=0.3\columnwidth,angle=0]{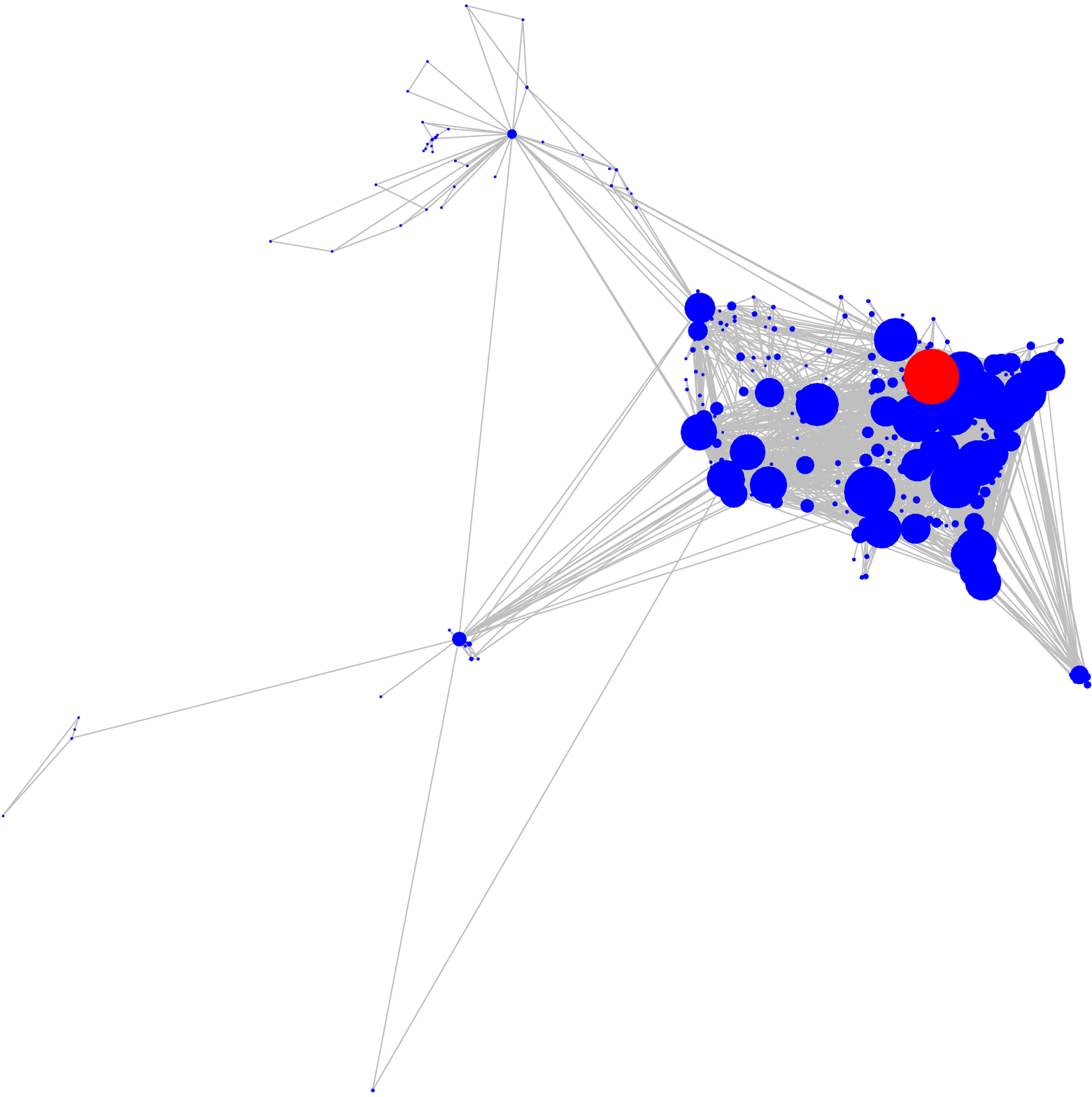}
b)\includegraphics[width=0.3\columnwidth,angle=0]{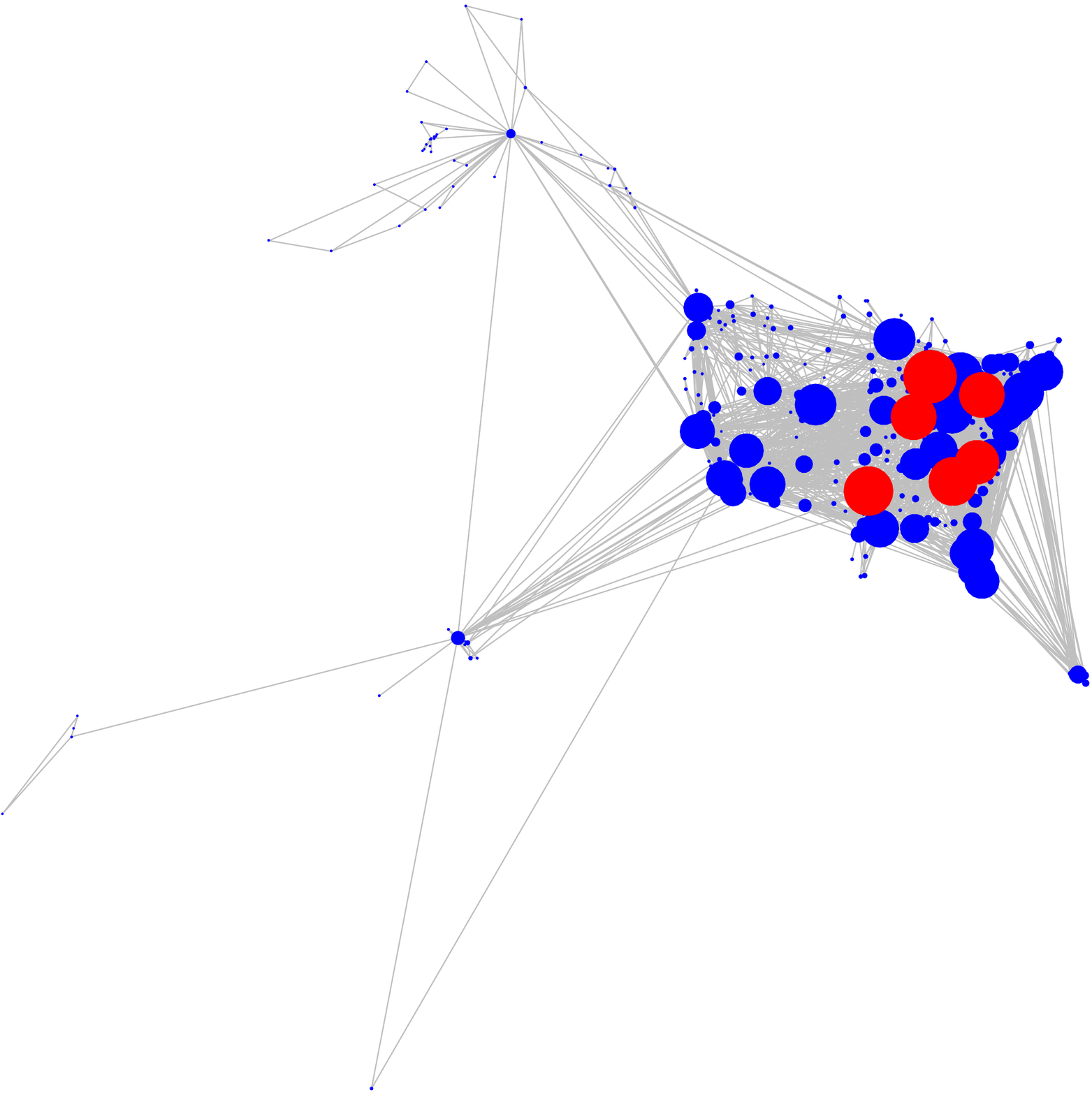}
c)\includegraphics[width=0.3\columnwidth,angle=0]{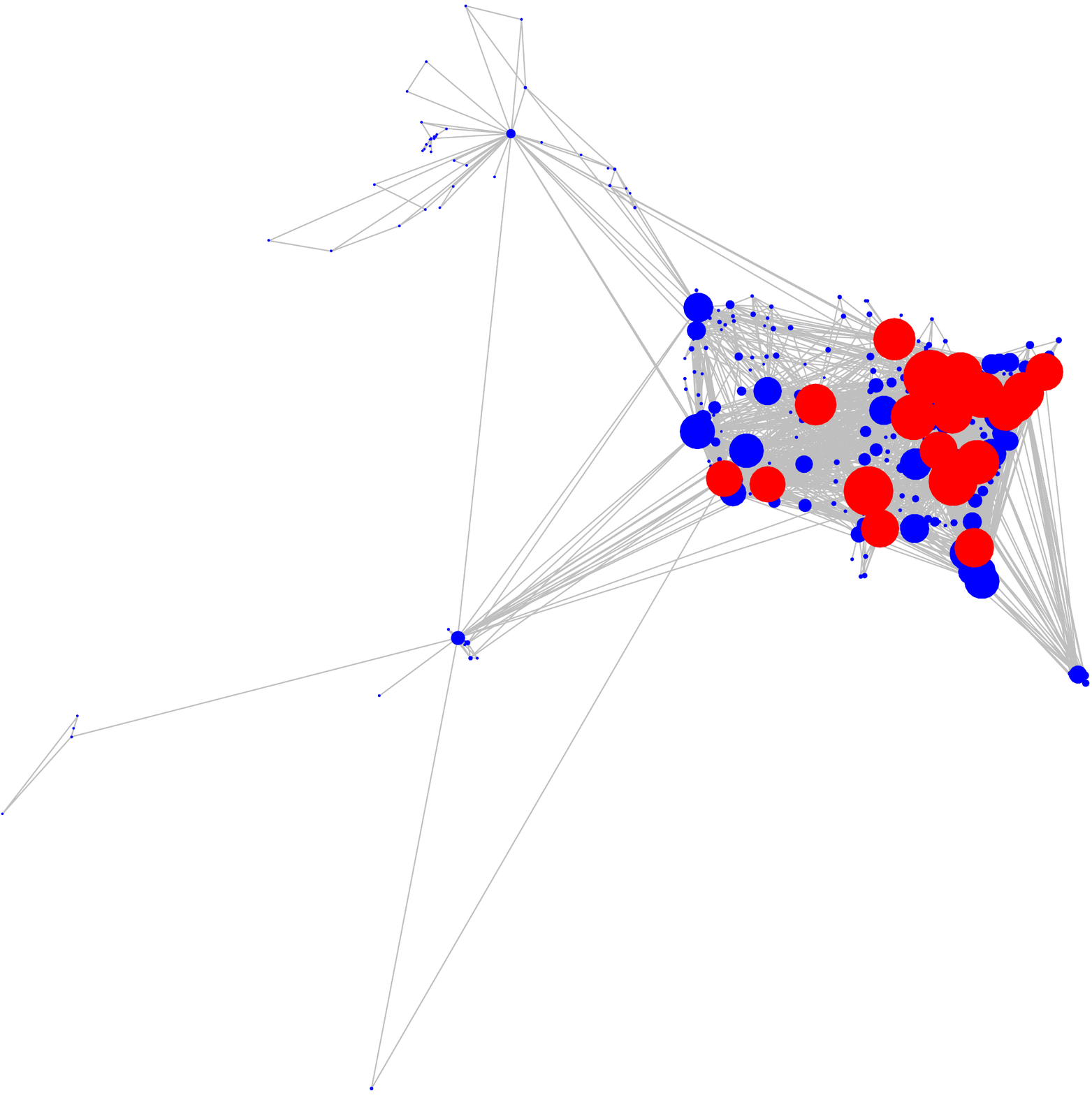}
%
%
%
%

\caption{Snapshots of the melting process of the USAir97 network at three different
values of $\beta$, namely at$1.5\cdot10^{-7}$(a),$1.25\cdot10^{-7}$(b),$1.0\cdot10^{-7}$(c). }

\label{melting_USAir}
\end{figure}

This result has important consequences for the robustness of networks.
Those networks displaying a high robustness to external stresses,
such that $\beta_{c}$ is very close to zero, start their melting
by the most central nodes according to EC. That is, if we consider
a network like the USA transportation network, which has $\beta_{c}$
of the order of $10^{-7},$ we will observe that the first airports
to be disconnected from the giant connected component are the most
important ones in terms of their connectivity. Here we give the list
of the first airports separated from the giant connected component
in order of their disconnection: Chicago O'Hare, Dallas/Forth Worth
Int., The William B. Hartsfield (Atlanta), Detroit Metropolitan, Pittsburgh
Intel., Lambert-St. Louis, Charlotte/Douglas Int. (see Fig. \ref{melting_USAir}).

\section{Conclusions and future outlook}

The most important result of the current work is the proof of the
existence of a universal melting transition in graphs and networks.
This transition takes place when we consider a Lindemann-like model
on graphs, which is based on a vibrational approach of nodes and edges.
From a mathematical point of view the current method is based on the
spectral properties of the adjacency matrix of the graph and the changes
taking place on the exponential matrix function $\exp\left(\beta A\right)$
with the changes of $\beta$. In this way, regular-like graphs like
square lattices are easier to melt than more irregular structures,
such as spatial planar graphs. These differences resemble the known
dissimilarities between crystalline and amorphous solids in their
melting. 

The analysis of a series of real-world networks has given us the possibility
of exploring the global and local structural characteristics of networks
which drives their melting. At the global topological level, we have
shown here that the value $\beta_{c}$ at which the melting of a graph
occurs depends mainly on the differences between the local and global
degree heterogeneities existing in the graph. At the local one we
have observed that the melting is triggered by the nodes having the
higher eigenvector centrality in the network, particularly in those
cases where the melting temperature is very close to zero. 

The analysis of graph/networks melting as proposed here opens many
new possibilities for the study of network robustness to external
stresses. There are many mathematical and computational questions
that remain open from the current study. They include, but are not
limited, to the following ones: (i) A more exhaustive analysis of
the topological (global and local) drivers of the graph melting; (ii)
How certain specific network characteristics, e.g., clustering, modularity,
degree assortativity, etc., influence the melting temperature and
the melting process of artificial and real-world networks?; (iii)
Is there a ranking of certain classes of graphs, e.g., trees, monocyclic
graphs, etc., according to their melting? We hope the reader can help
to answer some of these questions and generate new ones that clarify
our understanding of network robustness to external stresses.

\section*{Acknowledgement}

The authors thank Dr. F. Arrigo, and Prof. D. H. Higham for useful
comments and suggestions which improve the presentation of the material.
NA thanks Iraqui Government for a Doctoral Fellowship at the University
of Strathclyde.

\bibliographystyle{siam}

\end{document}